\RequirePackage{rotating}
\documentclass[11pt,a4paper,reqno]{amsart}

\usepackage[utf8]{inputenc}
\usepackage[english]{babel}
\usepackage{amsmath,threeparttable}
\usepackage{subfig}
\usepackage{amssymb, mathabx}
\usepackage[numbers]{natbib}
\usepackage{graphicx}
\usepackage{braket}
\usepackage{float}
\restylefloat{table}
\usepackage{booktabs}
\usepackage{rotating}
\usepackage{enumerate}
\usepackage[dvipsnames]{xcolor}
\usepackage{framed}
\usepackage[labelfont=rm]{caption}
\usepackage{multirow}
\usepackage{siunitx}
\usepackage{longtable}
\usepackage[colorlinks,hyperindex,bookmarksopen,linkcolor=black,citecolor=black,urlcolor=black]{hyperref}

\usepackage{mathrsfs}

\usepackage{braket}

\bibpunct{[}{]}{;}{n}{,}{,}
\usepackage[hmargin=3cm,vmargin={3.5cm,4cm}]{geometry}

\theoremstyle{definition}                                 

\theoremstyle{definition}                           

\theoremstyle{remark}                             

\usepackage{color}

\usepackage{mathtools,slashed}

\newcommand{\be}{\begin{eqnarray}}
\newcommand{\ee}{\end{eqnarray}}

\allowdisplaybreaks

\begin{document}
\title[A modeling and simulation study of AD in plug-flow reactors]{A modeling and simulation study of anaerobic digestion in plug-flow reactors}

\author{DB Panaro
  \and
  MR Mattei
  \and
  G Esposito
  \and
  JP Steyer
  \and
  F Capone
  \and
  L Frunzo
}

\newcommand{\Addresses}{{
  \bigskip
  \footnotesize

  DB Panaro, \textsc{Department of Mathematics and applications "R. Caccioppoli", University of Naples Federico II, via Cintia, Monte S.Angelo, 80126 Naples, Italy}\par\nopagebreak
  \textit{E-mail address}, DB Panaro: \texttt{danielebernardo.panaro@unina.it}

  \medskip

  MR Mattei, \textsc{Department of Mathematics and applications "R. Caccioppoli", University of Naples Federico II, via Cintia, Monte S.Angelo, 80126 Naples, Italy}\par\nopagebreak
  \textit{E-mail address}, MR Mattei: \texttt{mariarosaria.mattei@unina.it}

  \medskip

  G Esposito, \textsc{Department of Civil, Architectural and Environmental Engineering, University of Naples Federico II, via Claudio 21, 80125 Naples, Italy}\par\nopagebreak
  \textit{E-mail address}, G Esposito: \texttt{gioespos@unina.it}
  
    \medskip

  JP Steyer, \textsc{LBE, Univ Montpellier, INRAE, 102 avenue des Etangs, 11100 Narbonne, France}\par\nopagebreak
  \textit{E-mail address}, JP Steyer: \texttt{Jean-Philippe.Steyer@inrae.fr}
  
      \medskip

  F Capone, \textsc{Department of Mathematics and applications "R. Caccioppoli", University of Naples Federico II, via Cintia, Monte S.Angelo, 80126 Naples, Italy}\par\nopagebreak
  \textit{E-mail address}, F Capone: \texttt{fcapone@unina.it}
  
        \medskip

  L Frunzo, \textsc{Department of Mathematics and applications "R. Caccioppoli", University of Naples Federico II, via Cintia, Monte S.Angelo, 80126 Naples, Italy}\par\nopagebreak
  \textit{E-mail address}, L Frunzo: \texttt{luigi.frunzo@unina.it}

}}

%
%



\begin{abstract}

In this work a mathematical model for the anaerobic digestion process in plug-flow reactors is proposed on the basis of mass balance considerations. The model consists of a system of parabolic partial differential equations for the variables representing the concentrations of the bio-components constituting the waste matrix and takes into account convective and diffusive phenomena. The plug-flow reactor is modelled as a one-dimensional domain; the waste matrix moves in the direction of the reactor axis and undergoes diffusive phenomena which reproduce the movement of the bio-components along the reactor axis due to a gradient in concentration. The velocity characterizing the convection of the waste matrix moving within the reactor is not fixed a priori but it is considered as an additional unknown of the mathematical problem. The variation in the convective velocity allows to account the mass variation occurring along a plug-flow reactor due to the conversion of solids, which is an aspect not much analysed in the literature of dry anaerobic digestion in plug-flow reactors. The equation governing the convective velocity is derived by considering the following hypothesis: the density of the waste matrix within the reactor is supposed constant over time and the sum of the volume fractions of the bio-components constituting the waste matrix are constrained to sum up to unity. The waste matrix undergoes biochemical transformations catalysed by anaerobic microbial species which lead to the production of gaseous methane, the final product of the anaerobic digestion process. Biochemical processes are modelled using a simplified scheme and a differential equation is used to describe the dynamics of the produced gaseous methane. A finite difference scheme is used for the numerical integration. Lastly, the model consistency is showed through numerical simulations which investigate the effect of the variation of some operating parameters on the process performance. The model is then applied to a real case scenario of engineering interest. Simulations produce results in agreement with the experimental observations. This highlights that the model can serve as a tool for the optimal management and sizing of an anaerobic digestion plug-flow reactor.

\end{abstract}

\maketitle

\section{Introduction} \label{n1}
Nowadays, Anaerobic Digestion (AD) process is a technology widely used for the treatment of the Organic Fraction of Municipal Solid Waste (OFMSW). The organic compounds are converted in biogas thanks to the activity of various microbial groups operating in an oxygen-free environment. The biogas, due to its high methane content, can be used to produce simultaneously heat and electricity through combined heat and power systems or directly injected in the gas grid (after cleaning) and is considered as a renewable energy. The residual substrate of the AD represents a by-product of the process, called digestate, which is used as fertilizer in agriculture.

The AD process is usually classified based on the Total Solids (TS) content of the substrate to be treated, where TS is the measure of all the suspended, colloidal and dissolved solids in a medium. If the TS content is less or equal to 10\% the AD is classified as Wet-AD (WAD) otherwise, if it is greater or equal to 20\%, it is denoted as Dry-AD (DAD). Lastly, the AD is denoted as semi-dry AD (SDAD) when the TS content is between 10\% and 20\%. Moisture content is essential to the AD process because it dissolves nutrients and facilitates the contact between substrates and bacteria \cite{le2012influence}. Compared to WAD, DAD has the advantage to reduce the needed reactor volume, the energy needs as there is less water to heat up and, moreover, the amount of by-product to be treated  \cite{xu2015mathematical, bolzonella2006dry}. However, DAD usually presents some drawbacks related to inhibition, accumulation of Volatile Fatty Acids (VFA) and long Hydraulic Retention Time (HRT) \cite{kothari2014different}. VFAs are short chain carboxylic acids having from 2 to 5 carbon atoms in the molecule and are among the essential intermediates of the AD process. HRT indicates the mean residence time of a certain substrate within a biological
reactor \cite{reif2013removal}, determining the contact time between the substrate to be treated and the microorganisms. One of the  most used reactor configuration for DAD processes is the Plug-Flow Reactor (PFR), constituted by a reactor having a tubular or parallelepiped shape. In a plug-flow reactor, the waste movement along the digester is such that back-mixing is avoided \cite{srinivasan2004plug} and a portion of the effluent is typically recycled to inoculate the influent and improve the AD process \cite{li2014spatial}. The hydrodynamics of a full-scale plug-flow reactor depends on the adopted configuration. For example Dranco systems work as vertical downflow digesters, where the mixing occurs via recirculation of the waste extracted at the bottom of the reactor; an horizontal flow is used in the Kompogas configuration, realized through slow rotation mixers that ensure the homogenization in radial direction and the stripping of the biogas; the horizontal plug-flow is circular in the Valorga configuration, where the mixing is realized injecting the biogas from the bottom of the reactor \cite{lissens2001solid}.
\\
Mathematical modeling of AD process represents a powerful tool to enhance process control and optimization together with reactor sizing and plant management. The vast majority of models describing WAD process refers to Continuous Stirred Tank Reactor (CSTR) configuration and follows the approach of the Anaerobic Digestion Model No.1 (ADM1) \cite{batstone2002iwa, frunzo2019adm1, maharaj2019adm1}. ADM1 consists of a system of Ordinary Differential Equations (ODEs) which simulate the main biochemical processes occurring during anaerobic digestion. The latter can be classified into five main phases: disintegration, hydrolysis, acidogenesis, acetogenesis and methanogenesis. According to \cite{batstone2013modelling}, one of the main limitations to mathematical modeling of AD is related to the lack of reliable mathematical models capable to describe this kind of process in PFRs. Contrary to CSTR, where the hypothesis of complete mixing of the compounds inside the reactor is considered, in a PFR model the state variables are functions both of time and space. Moreover, perfect mixing in radial direction is assumed, then reactant concentrations are uniform in any cross section and vary only along the flow path. The system of PDEs describing the phenomenon, is represented by mass balances on the state variables of the problem. In literature there exist a few number of works related to the modelling of AD in PFRs due to the uncertainty linked to some aspects of the problem such as dispersion, mixing, turbulence, variation of density and variation of the porosity of the medium. The existing PFR models make some simplifications to describe the phenomenon. For example in \cite{donoso2018modelling} the authors approximate the PFR as a tank-in-series system, solving a sequence of CSTR-type reactors; in \cite{vavilin2007anaerobic} Vavilin et al. solved the convection-diffusion-reaction equation for a PFR using a constant convective velocity; Binxin Wu in \cite{wu2012integration} faced the problem using a two-stage computational fluid dynamics (CFD) where the physical model, consisting of a system of PDEs, is solved under steady-state conditions while the biological model is represented by a system of ODEs.

Other works focus on the discussion of the boundary conditions for the convection-diffusion-reaction equations describing the PFR configuration. They mostly debate on the validity of the boundary conditions postulated by P.V. Danckwerts in 1953 \cite{danckwerts1953continuous}. In his work Danckwerts argues that for a convection-diffusion-reaction equation at the inlet section the flux continuity must be preserved while at the outlet the gradient in concentration must be zero. Several authors successfully used the Danckwerts' boundary conditions in their works \cite{bischoff1961note, pearson1959note, van1966further, deckwer1976boundary}, others criticized some aspect of Danckwerts' assertions, mostly claiming that the state variables should preserve the continuity across the inlet boundary \cite{wicke1975physical} or that the mass and heat flux are not conserved if the contribution of dispersion to axial mass flow is important \cite{delgado2008analysis}.\\
The aim of this work is to present a mathematical model capable to describe the DAD process organic waste in a PFR. The model considers the process being governed by one-dimensional convection-diffusion-reaction equations where the convective velocity is not a fixed value but it is considered as a function both of space and time. In the existing models, the general approach consists in neglecting the system mass variation due to the degradation processes that lead to the conversion of solids in gaseous compounds along the reactor. 
In this new kind of approach it is considered that this loss in mass is balanced by the variation of the convective velocity of the system, taking into account that these kind of processes are performed maximizing the working reactor volume, keeping constant the level of the treated substrate along the reactor. The equation governing the system velocity variation is derived through the hypothesis that the waste density is constant in time and space and the sum of the volume fractions of the bio-components constituting the waste matrix are constrained to sum up to unity. These statements imply that the mass of the waste mixture should remain constant in time and space, hence the solids mass reduction due to the degradation processes along the reactor, as aforementioned, has to be balanced by the variation of the convective velocity. Suitable initial-boundary conditions are prescribed. 
Biochemical processes are modelled using a simplified kinetic scheme where disintegration and methanogenesis are considered as the rate-limiting steps of the whole digestion process. Equations are numerically integrated by using a finite difference scheme and numerical simulations show different aspects that could be analysed  during the management of an existing plant or the designing of new plants. They investigate the roles of some operating and physical parameters on process dynamics in terms of methane production, removal efficiency, bio-components constituting the waste matrix concentration trends and convective velocity trend. The investigated physical and operating parameters are the reactor length, the inlet convective velocity of the waste, the HRT and the diffusion coefficient. Finally, the model is applied to the case of a lab-scale experiment reproducing the DAD process in a PFR for different Organic Loading Rates (OLRs), where the OLR indicates a measure of the amount of organic material per unit reactor volume subjected to an AD process in a given unit time period \cite{grangeiro2019new}.

\section{Mathematical model} \label{n2}

 \subsection{Model equations} \label{n2.1}
\begin{figure}
\centering
\fbox{\includegraphics[scale=0.055]{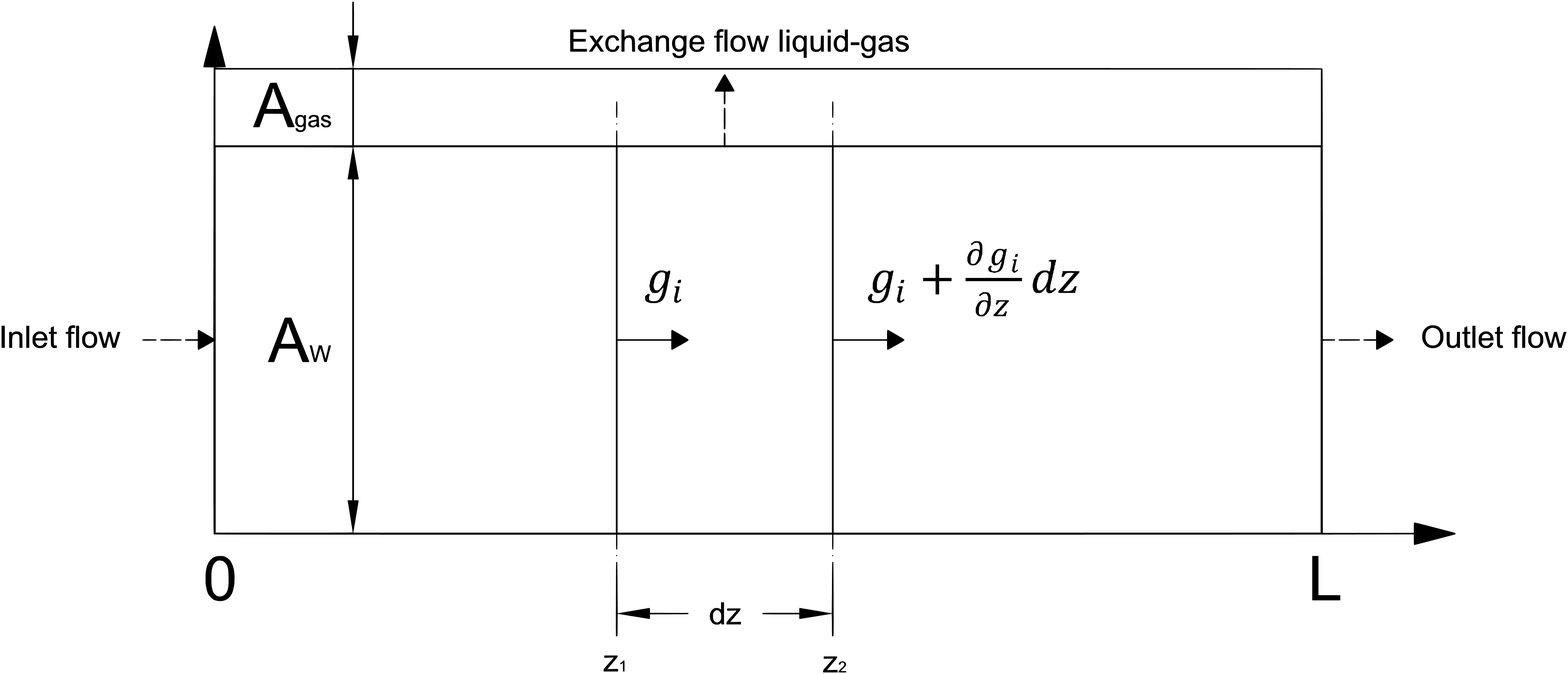}}
\caption{Control volume for the mass balance.} \label{fig1}     
\end{figure}
The mathematical model is derived in the framework of continuum mechanics and is based on mass balance considerations in one-dimensional case for $n$ state variables $C_i\left(z,t\right), \ i=1,...,n, \ \textbf{C}=\left(C_1,...,C_n\right)$ which represent the concentrations of the bio-components constituting the waste mixture. The state variables are considered as functions of time $t$ and space $z$, where $z$ represents the spatial coordinate oriented along the reactor axis and directed from the inlet to the outlet section. Let us consider a control volume $A_{w}dz$ as represented in Figure \ref{fig1}, where $A_{w}$ is the constant cross-sectional area of the reactor occupied by the waste.
Taking into account the mass flux per unit area $g_i\left(z,t\right)$ crossing the surface $A_{w}$ and the source/consumption term $F_i\left(z,t,\textbf{C}\right)$, the mass balance on the $i^{th}$ compound gives:
\setcounter{equation}{0}
\begin{multline}
A_w\frac{\partial}{\partial t}\int_{z_1}^{z_2} C_i\left(z,t\right)\ dz =A_{w}\left[g_i\left(z_1,t\right)-g_i\left(z_2,t\right)\right] + A_w\int_{z_1}^{z_2} F_i\left(z,t,\textbf{C}\right)\ dz, \\
 0< z< L,\quad t>0,\quad i=1,...,n, \label{eq1}
\end{multline}
\begin{multline}
\int_{z_1}^{z_2} \frac{\partial C_i\left(z,t\right)}{\partial t}\ dz = -\int_{z_1}^{z_2}\frac{\partial g_i\left(z,t\right)}{\partial z}\ dz + \int_{z_1}^{z_2} F_i\left(z,t,\textbf{C}\right)\ dz ,\\ 0< z< L,\quad t>0,\quad i=1,...,n. \label{eq2}
\end{multline}

where:
\begin{itemize}
\item[-] $g_i\left(z,t\right) = v_i\left(z,t\right)C_i\left(z,t\right) - \bar{D_i}\left(z,t\right)\partial C_i(z,t)/ \partial z, \ i=1,...,n$ ;
\item[-] $F_i\left(z,t,\textbf{C}\right)=	\sum_{j=1}^k \alpha_{ij}r_j\left(z,t,\textbf{C}\right), \ i=1,...,n$;
\item[-] $k$ is the number of processes involved;
\item[-] $\alpha_{ij}$ is the stoichiometric coefficient of compound $i$ on process $j$;
\item[-] $r_j\left(z,t,\textbf{C}\right)$ is the kinetic rate of process $j, \ j=1,...,k$.
\end{itemize} 
The convective flux of each compound of the waste matrix is characterized by a velocity $v_i\left(z,t\right)$ which is supposed to be a function of both space and time but the same for each compound. The diffusion coefficient $\bar{D_i}$ is assumed constant along the z-direction and on time and equal for all the compounds:
\begin{equation}
v_i=v, \quad \bar{D_i}=\bar{D}, \quad i=1,...,n. \label{eq4}
\end{equation}
Differentiating equation \eqref{eq2} with respect to $z_2$ and considering $z_2=z$ follows:
\begin{multline}
\frac{\partial C_i\left(z,t\right)}{\partial t} + \frac{\partial \left(v\left(z,t\right)C_i\left(z,t\right)\right)}{\partial z} - \bar{D} \frac{\partial^2C_i\left(z,t\right)}{\partial z^2}=F_i\left(z,t,\textbf{C}\right), \\ 0<z< L,\quad t> 0,\quad i=1,...,n. \label{eq5}
\end{multline}
Equation \eqref{eq5} represents the well known convection-diffusion-reaction equation in conservative form in one-dimensional case. 

Now let us consider the composition of the waste matrix moving along the reactor. This matrix has a density $\rho\left(z,t\right)=\rho$ supposed to be constant with space and time and is constituted by water, particulate and dissolved components. The particulate fraction is composed of inerts, volatile solids (VS) (divided in bio-degradable and non bio-degradable material) and microbial biomass. VS represents the portion of the TS content that is volatilized at $\SI{550}{\celsius}$ and gives an idea on the amount of the readily vaporizing matter present in the solid fraction of a substrate. The sum of inerts and VS constitute the TS content of the waste. The dissolved components are soluble acetic acid and soluble methane, which are anaerobically produced from the degradation of VS performed by the microbial biomass. Their concentrations are usually expressed in terms of Chemical Oxygen Demand (COD) which is a measure of the amount of oxygen that is needed for the complete chemical oxidation of organic compounds of a medium. It is commonly expressed in mass of oxygen consumed over volume of the medium sample used for its measurement. 

The volume of the head-space above the waste matrix is considered constant and the gas occupying this volume is composed only by gaseous methane, whose concentration is considered invariable along the space.\\

The following notations will be used:
\begin{enumerate}[-]
\item $X_1\left(z,t\right)$ is the water concentration within the reactor;
\item $X_2\left(z,t\right)$ is the inerts concentration within the reactor; 
\item $X_3\left(z,t\right)$ is the concentration of bio-degradable VS;
\item $X_4\left(z,t\right)$ is the concentration of non bio-degradable VS;
\item $X_5\left(z,t\right)$ is the microbial biomass concentration within the reactor expressed in terms of VS; 
\item $S_1\left(z,t\right)$ is the soluble acetic acid concentration within the reactor expressed in terms of COD;
\item $S_2\left(z,t\right)$ is the soluble methane concentration within the reactor expressed in terms of COD;
\item $G\left(t\right)$ is the gaseous methane concentration within the head-space of the reactor expressed in terms of COD;
\item $\textbf{X}=\left(X_1,...,X_5\right)$;
\item $\textbf{S}=\left(S_1,S_2\right)$;
\item $F_{X,h}\left(z,t,\textbf{X},\textbf{S},G\right), \ h=1,...,5,$ is the source/consumption term of the particulate compound $X_h$;
\item $F_{S,l}\left(z,t,\textbf{X},\textbf{S},G\right), \ l=1,2,$ is the source/consumption term of the dissolved compound $S_l$.
\\
\end{enumerate}

\begin{figure}
\centering
\fbox{\includegraphics[width=0.5\textwidth, keepaspectratio]{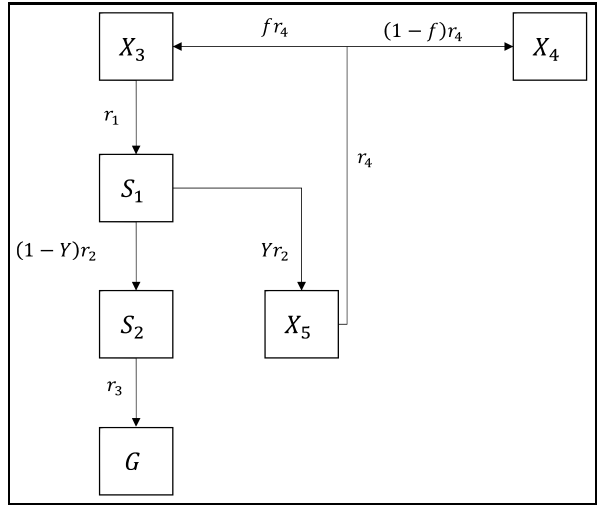}}
\caption{Kinetic scheme.} \label{fig2}     
\end{figure}

The kinetic scheme of the model is reported in the Figure \ref{fig2}. The conversion of the bio-degradable VS $X_3$ in soluble acetic acid $S_1$, which takes place through different processes that are disintegration, hydrolysis, acidogenesis and acetogenesis, is summarized in a unique kinetic rate $r_1$, considering only the rate-limiting step that is the disintegration during AD of complex substrates \cite{esposito2011model, fiore2016scale, van2008anaerobic}; from the degradation of the acetic acid through the non linear Monod-type kinetic rate $r_2$ are produced soluble methane $S_2$ and microbial biomass $X_5$; the microbial biomass die according to the decay law whose kinetic rate is  indicated as $r_4$ and produce new bio-degradable and non-biodegradable VS $X_4$; lastly, the soluble and gaseous methane ($S_2$ and $G$), are in equilibrium according to the gas-transfer law expressed by the kinetic rate $r_3$.\\

Applying equation (\ref{eq5}) to the compounds $X_h, \ h=1,...,5$, $S_1$ and $S_2$ the following non-linear system of PDEs is obtained.

\begin{multline}
\frac{\partial X_h(z,t)}{\partial t} + \frac{\partial (v(z,t)X_h(z,t))}{\partial z} - \bar{D} \frac{\partial^{2} X_h(z,t)}{\partial z^{2}}=F_{X,h}\left(z,t,\textbf{X},\textbf{S},G\right), \\ 0< z< L,\quad t>0,\quad h=1,...,5, \label{eq6}
\end{multline}
\begin{multline}
\frac{\partial S_l(z,t)}{\partial t} + \frac{\partial (v(z,t)S_l(z,t))}{\partial z} - \bar{D} \frac{\partial^{2} S_l(z,t)}{\partial z^{2}}=F_{S,l}\left(z,t,\textbf{X},\textbf{S},G\right), \\ 0< z< L,\quad t>0, \quad l=1,2, \label{eq7} 
\end{multline}

where:
\begin{itemize}
\item $F_{X,1}\left(z,t,\textbf{X},\textbf{S},G\right)=F_{X,2}\left(z,t,\textbf{X},\textbf{S},G\right)=0$;
\item $F_{X,3}\left(z,t,\textbf{X},\textbf{S},G\right)=fr_4 -r_1$;
\item $F_{X,4}\left(z,t,\textbf{X},\textbf{S},G\right)= (1-f)r_4$;
\item $F_{X,5}\left(z,t,\textbf{X},\textbf{S},G\right)=Yr_2-r_4$;
\item $F_{S,1}\left(z,t,\textbf{X},\textbf{S},G\right)= m(r_1 -r_2)$;
\item $F_{S,2}\left(z,t,\textbf{X},\textbf{S},G\right)=m(1-Y)r_2-r_3$;
\item $r_1=k_1 X_{3}$;
\item $r_2=k_2 X_{5}S_{1}/\left(K_{1}+S_{1}\right)$;
\item $r_3=k_3(S_{2}-RTK_HG)$;
\item $r_4=k_4X_{5}$;
\item $m$ is the conversion factor of VS in COD $ \left[g_{COD} \ g_{VS}^{-1}\right]$;
\item $Y$ is the yield of biomass on substrate;
\item $f$ is the fraction of dead microbial biomass becoming new bio-degradable substrate;
\item $k_1$ is the kinetic constant for the consumption of the volatile solids $X_3$, having the dimension of $\left[T^{-1}\right]$;
\item $k_{2}$ is the Monod maximum specific uptake rate for the acetic acid $\left[T^{-1}\right]$;
\item $K_{1}$ is the half saturation constant $\left[M \ L^{-3}\right]$ for the kinetics of consumption of the acetic acid;
\item $k_3$ is the gas-liquid transfer coefficient $\left[T^{-1}\right]$;
\item $R$ is the gas law constant $\left[ L^2 \ T^{-2} \ \Theta^{-1}\right]$;
\item $T$ is the operating temperature $\left[\Theta\right]$;
\item $K_H$ is the Henry's law coefficient $\left[ L^2 \ T^{-2} \right]$;
\item $k_4$ is the first order decay rate of the microbial biomass $X_6$ $\ \left[T^{-1}\right]$. 
\\
\end{itemize}  

The velocity displacement of the waste matrix along the reactor axis constitutes an additional unknown of the problem, and a further equation is needed to describe its variation in space and time. To this aim, the following hypothesis is made: the sum of the volume fractions of the particulate components within the reactor is constrained to sum up to unity, that is:

\begin{equation}
   \begin{cases}
   \sum_{h=1}^5 X_h\left(z,t\right)/\rho_h=1,\\ 
   \rho_h=\rho, \quad h=1,...,5
   \end{cases} \Longrightarrow \quad \sum_{h=1}^5 X_h (z,t)=\rho. \label{eq8}
\end{equation}

Equation \eqref{eq8} implies that the mass of the mixture composed of water, inerts, VS and microbial biomass is constant over time. As a consequence, the convective velocity of the waste matrix varies along the reactor and its variation depends on the kinetics of the compounds constituting the mixture. In fact, the velocity variation has to balance the consumption of the VS to keep the mass of this particular mixture constant. 

These considerations are used to derive the equation governing the velocity. Summing equation \eqref{eq6} on $h=1,...,5$ and taking into account equation \eqref{eq8} follows:

\begin{align}
&\frac{\partial v(z,t)}{\partial z}=\frac{\sum_{h=1}^5 F_{X,h}\left(z,t,\textbf{X},\textbf{S},G\right)}{\rho}, \label{eq9} \\
&\sum_{h=1}^5 F_{X,h}\left(z,t,\textbf{X},\textbf{S},G\right)=Yr_2-r_1. \label{eq9.1}
\end{align}

In addition, a differential equation is derived from the mass balance for the volume $V_{gas}=A_{gas}L$ of the head-space where the biogas is stored. The dynamics of the gaseous methane $G\left(t\right)$ is described (Eq. \eqref{eq10}):

\begin{equation}
\frac{dG(t)}{dt}=\frac{A_{w}}{V_{gas}}\int_{0}^{L}r_3\left(z,t\right)\ dz. \label{eq10}
\end{equation}

This equation describes the fact that all the contributes to the gas-transfer in each point are summed to define a unique gas-transfer rate. The ratio between the cross-section occupied by the waste and the volume of gas is present to take into account the fact that the gas-transfer kinetic rate is waste volume-specific.

\subsection{Boundary and initial conditions} \label{n2.2}
With the aim to set a closed problem, initial-boundary conditions are prescribed.
Firstly, the convective velocity of compounds moving along the reactor at the boundary $z=0$ is assumed equal to the incoming flow velocity (Eq. \eqref{eq11}):

\begin{equation}
v(0,t)=v_0,\quad t\geq 0.
\label{eq11}
\end{equation}
\\
The value $v_0$ can be obtained by fixing the HRT of waste moving along the reactor of length $L$: 
\begin{equation}
v_0=\frac{L}{HRT}. \label{eq12}
\end{equation}
\\
For the PDEs \eqref{eq6} and \eqref{eq7} Danckwerts' boundary conditions \citep{danckwerts1953continuous}, which are Robin (Eqs. \eqref{eq13} and \eqref{eq15}) and Neumann (Eqs. \eqref{eq14} and \eqref{eq16}) conditions, are used. These conditions allow to preserve flux continuity in the first and last section of the reactor. Particularly, equations \eqref{eq13} and \eqref{eq15} indicate that, considering each particulate or dissolved compound constituting the mixture, the difference in concentration between the incoming flow rate and the first section of the reactor is linked to the diffusion phenomenon. Moreover, equations \eqref{eq14} and \eqref{eq16} imply that the diffusive flux has to be null in the last section.

\begin{equation}
- \bar{D} \frac{\partial X_h(0,t)}{\partial z}=v_0(X_{h,IN}-X_h(0,t)), \quad h=1,...,5, \quad t>0,
\label{eq13}
\end{equation}

\begin{equation}
\frac{\partial X_h(L,t)}{\partial z}=0, \quad h=1,...,5, \quad t>0,
\label{eq14}
\end{equation}

\begin{equation}
- \bar{D} \frac{\partial S_l(0,t)}{\partial z}=v_0(S_{l,IN}-S_l(0,t)), \quad l=1,2, \quad t>0,
\label{eq15}
\end{equation}

\begin{equation}
\frac{\partial S_l(L,t)}{\partial z}=0, \quad l=1,2, \quad t>0.
\label{eq16}
\end{equation}
\\

In \eqref{eq13} and \eqref{eq15} $X_{h,IN}$ and $S_{l,IN}$ are the concentrations of each particulate and dissolved compound in the incoming flow rate, respectively.

Lastly, the following initial conditions are considered (Eqs. \eqref{eq17}, \eqref{eq18} \eqref{eq19}):

\begin{equation}
X_h(z,0)=X_{h,0},\quad h=1,...,5,\quad 0\leq z \leq L, \label{eq17}
\end{equation}

\begin{equation}
S_l(z,0)=S_{l,0},\quad l=1,2 \quad 0\leq z \leq L, \label{eq18}
\end{equation}

\begin{equation}
G(0)=G_0. \label{eq19}
\end{equation}

\section{Numerical simulations} \label{n3}
\subsection{Model input} \label{n3.1}
The numerical method used for the integration of the system of PDEs stated in Section \ref{n2.1} with boundary and initial conditions presented in Section \ref{n2.2} follows the philosophy of the finite difference upwind method \cite{d2012computational}. The method is conditionally stable and was implemented through an original software developed using Matlab platform.\\
Four sets of simulations $A$, $B$, $C$ and $D$ are performed to show model consistency. The effects of some operating and physical parameters on process performance are investigated in terms of methane production, bio-degradable VS removal efficiency, acetic acid concentration trend, microbial biomass concentration trend and convective velocity trend. The results are shown for a specific simulation time or, for one set (set $C$), for different simulation times. The investigated physical and operating parameters are the reactor length $L$, the inlet convective velocity of the waste $v_0$, the $HRT$ and the diffusion coefficient $\bar{D}$.
In Table \ref{table1} are reported the kinetic parameters used in all simulations. 
The values of $k_1$ and $k_3$ were defined in this study. Particularly, the value of the gas-liquid transfer coefficient was decreased with respect to its value used in the ADM1 \cite{batstone2002iwa} to take into account the limitation to the gas-liquid transfer due to the high solids content \cite{manchala2017anaerobic, abbassi2012total, liu2016characterization}. 
Table \ref{table2} summarizes the initial and boundary conditions adopted for the various simulation sets. Such conditions have been obtained by fixing the TS content of the substrate to 20\% and considering that the overall density of the treated substrate is equal to the density of water. Moverover, the VS content on TS base has been set equal to 70\% and it is 50\% non bio-degradable. The bio-degradable VS and the microbial biomass constitute the remaining 50\% of the VS content. In all simulations the conversion coefficient of VS in COD has been set equal to $m=1.5 \  g_{COD} \ g_{VS}^{-1}$. The fraction of decayed microbial biomass becoming new bio-degradable substrate was set equal to $f=0.2$.
\begin{table}
\scriptsize
\begin{center}
\renewcommand\arraystretch{1.3}
\begin{tabular}{llccc}
\toprule
{\textbf{Parameter}} & {\textbf{Definition}} & {\textbf{Unit}} &  {\textbf{Value}} & {\textbf{Reference}}\\
\midrule
$k_{1}$   & \begin{tabular}[c]{@{}l@{}}\vspace{-0.1cm} Kinetic constant for the consumption \\of the volatile solids $X_3$ \end{tabular}  &  $d^{-1}$                              & 0.10 & This study \\
$k_{2}$   & \begin{tabular}[c]{@{}l@{}}\vspace{-0.1cm} Monod maximum specific uptake \\ rate for $S_{1}$ \end{tabular}               &  $d^{-1}$                              & 8.00   & \citep{batstone2002iwa}   \\
$k_3$    & Gas-liquid transfer coefficient  &  $d^{-1}$ &  20.0 & This study  \\
$k_{4}$         & \begin{tabular}[c]{@{}l@{}}\vspace{-0.1cm} First order decay rate of the \\ microbial biomass \end{tabular}  &  $d^{-1}$ & 0.02    & \citep{batstone2002iwa}      \\
$K_{1}$    & Half saturation constant  &  $g_{COD}\ l^{-1}$ & 0.15 & \citep{batstone2002iwa} \\
$Y$         & Yield of biomass on substrate  &  - & 0.05    & \citep{batstone2002iwa}      \\
$K_H$    & Henry's law coefficient  &  $M bar^{-1}$ & 0.0011 & \citep{batstone2002iwa} \\ 
\bottomrule
\end{tabular}
\caption{Kinetic parameters used in model simulations.} \label{table1}
\end{center}
\end{table}

\begin{table}
\begin{small}
\begin{center}
\renewcommand\arraystretch{1.3}
\begin{tabular}{llcccc}
\toprule
{\textbf{Parameter}} & {\textbf{Symbol}} & {\textbf{Unit}} &  \textbf{Value} \\
\midrule
Density of the waste &   $\rho$  &$g \ l^{-1}$    & 1000.0    \\
Initial $H_2O$ concentration &   $X_{1,0}$  &$g \ l^{-1}$    & 800.0    \\
Initial inert concentration  &  $X_{2,0}$ &  $g \ l^{-1}     $               & 60.0  \\
Initial bio-degradable VS concentration  &  $X_{3,0}$ &  $g_{VS} \ l^{-1}     $               & 69.3   \\
Initial non-bio-degradable VS concentration & $X_{4,0}$   & $g_{VS} \ l^{-1}     $               & 70.0   \\
Initial soluble acetic acid concentration & $S_{1,0}$ & $g_{COD} \ l^{-1}$                  & 0.0  \\
Initial soluble methane concentration  & $S_{2,0}$   & $g_{COD} \ l^{-1}       $             & 0.0  \\
Initial microbial biomass concentration&   $X_{5,0}$            &  $g_{VS} \ l^{-1}$                         & 0.7          \\
Initial gas-phase methane concentration     &      $G_0$              &  $g_{COD} \ l^{-1}$                         & 0.0 \\
Inlet $H_2O$ concentration &   $X_{1,IN}$  &$g \ l^{-1}$    & 800.0    \\
Inlet inert concentration  &  $X_{2,IN}$ &  $g \ l^{-1}     $               & 60.0   \\
Inlet bio-degradable VS concentration  &  $X_{3,IN}$ &  $g_{VS} \ l^{-1}     $               & 69.3   \\
Inlet non-bio-degradable VS concentration & $X_{4,IN}$   & $g_{VS} \ l^{-1}     $               & 70.0   \\
Inlet soluble acetic acid concentration & $S_{1,IN}$ & $g_{COD} \ l^{-1}$                  & 0.0  \\
Inlet soluble methane concentration  & $S_{2,IN}$         & $g_{COD} \ l^{-1}       $             & 0.0  \\
Inlet microbial biomass concentration&   $X_{5,IN}$              &  $g_{VS} \ l^{-1}$                         & 0.7 \\
\bottomrule
\end{tabular}
\caption{Initial and boundary conditions used in model simulations.} \label{table2}
\end{center}
\end{small}
\end{table}

Moreover, Table \ref{table:sets} present the values of the physical and operating parameters used in the simulation sets. The indicated HRT and OLR are referred to the inlet flow. In all simulations the cross section of the reactors is assumed equal to $\SI{1}{\meter^2}$ for simplicity. In the simulation set $A$, it is analysed the behaviour of the system when the process is performed using a constant HRT in reactors having different length and, as a consequence, a different inlet convective velocity of the waste $v_0$. In set $B$, a constant value is assumed for the inlet convective velocity of the waste $v_0$ while the reactor length and the HRT change. In set $C$ the reactor length is constant but the inlet convective velocity of the waste, strictly related to the OLR, and the HRT are variable. It has to be highlighted that, since the same inlet VS concentration is used in all the simulations, when the HRT changes, the OLR changes too. Lastly, the value of the diffusion coefficient $\bar{D}$ is considered constant among the simulation sets $A$, $B$ and $C$, while set D investigates the effects of the diffusion coefficient $\bar{D}$  on reactor performances.

\begin{sidewaystable}
\centering
\scalebox{0.85}{
\begin{tabular}{lcccccccccccccccc}
\toprule
\textbf{Parameter} & \textbf{Unit}                                                                                             & \multicolumn{4}{c}{\textbf{Set A}}            & \multicolumn{4}{c}{\textbf{Set B}}            & \multicolumn{4}{c}{\textbf{Set C}}            & \multicolumn{3}{c}{\textbf{Set D}} \\
                   & \multicolumn{1}{l}{}                                                                                      & $A$-1 & $A$-2 & $A$-3 & $A$-4 & $B$-1 & $B$-2 & $B$-3 & $B$-4 & $C$-1 & $C$-2 & $C$-3 & $C$-4 & $D$-1  & $D$-2 & $D$-3 \\ \midrule
$\bar{D}$                & $10^{-7} \ m^2 \ s^{-1}$ & 1.0       & 1.0       & 1.0       & 1.0       & 1.0       & 1.0       & 1.0       & 1.0       & 1.0       & 1.0       & 1.0       & 1.0       & $10^2$     & $10^3$    & $10^4$    \\
$v_0$              & $cm \ d^{-1}$                                                                                             & 8.33      & 16.67     & 25.00     & 33.33     & 33.33     & 33.33     & 33.33     & 33.33     & 20.0      & 25.0      & 33.33     & 50.0      & 33.33      & 33.33     & 33.33     \\
L                  & $m$                                                                                                       & 2.5       & 5.0       & 7.5       & 10.0      & 2.5       & 5.0       & 7.5       & 10.0      & 10.0      & 10.0      & 10.0      & 10.0      & 10.0       & 10.0      & 10.0      \\
HRT                & $d$                                                                                                       & 30.0      & 30.0      & 30.0      & 30.0      & 7.5       & 15.0      & 22.5      & 30.0      & 50.0      & 40.0      & 30.0      & 20.0      & 30.0       & 30.0      & 30.0      \\
OLR                & $g_{VS}\ l^{-1}\ d^{-1}$                                                                                  & 4.7       & 4.7       & 4.7       & 4.7       & 18.7      & 9.3       & 6.2       & 4.7       & 2.8       & 3.5       & 4.7       & 7.0       & 4.7        & 4.7       & 4.7      
\\
\bottomrule
\end{tabular}}
\caption{Physical and operating parameters used during the four simulation sets $A$, $B$, $C$ and $D$. $\bar{D}=$Diffusion coefficient, $v_0=$Inlet flow rate velocity, $L=$Reactor length, $HRT=$Hydraulic Retention Time and $OLR=$Organic Loading Rate.}
\label{table:sets}
\end{sidewaystable}

\subsection{Numerical results and discussion} \label{n3.2}
For each simulation set, the results are reported in terms of the bio-degradable VS concentration $X_{3}\left(z,\tau\right)$, acetic acid concentration $S_{1}\left(z,\tau\right)$, microbial biomass concentration $X_{5}\left(z,\tau\right)$ trends and $v\left(z,\tau\right)$ profile after $\tau=60$ days of simulation time. Furthermore, the methane production over time is shown. These trends are reported in Figures \ref{fig3}, \ref{fig4} and \ref{fig5} for simulation sets $A$, $B$ and $C$ respectively and in Figure \ref{fig10} for set $D$. 

\subsubsection{Set A} \label{3.2.1}
\begin{figure}
\centering
{\fbox{\includegraphics[width=0.95\textwidth, keepaspectratio]{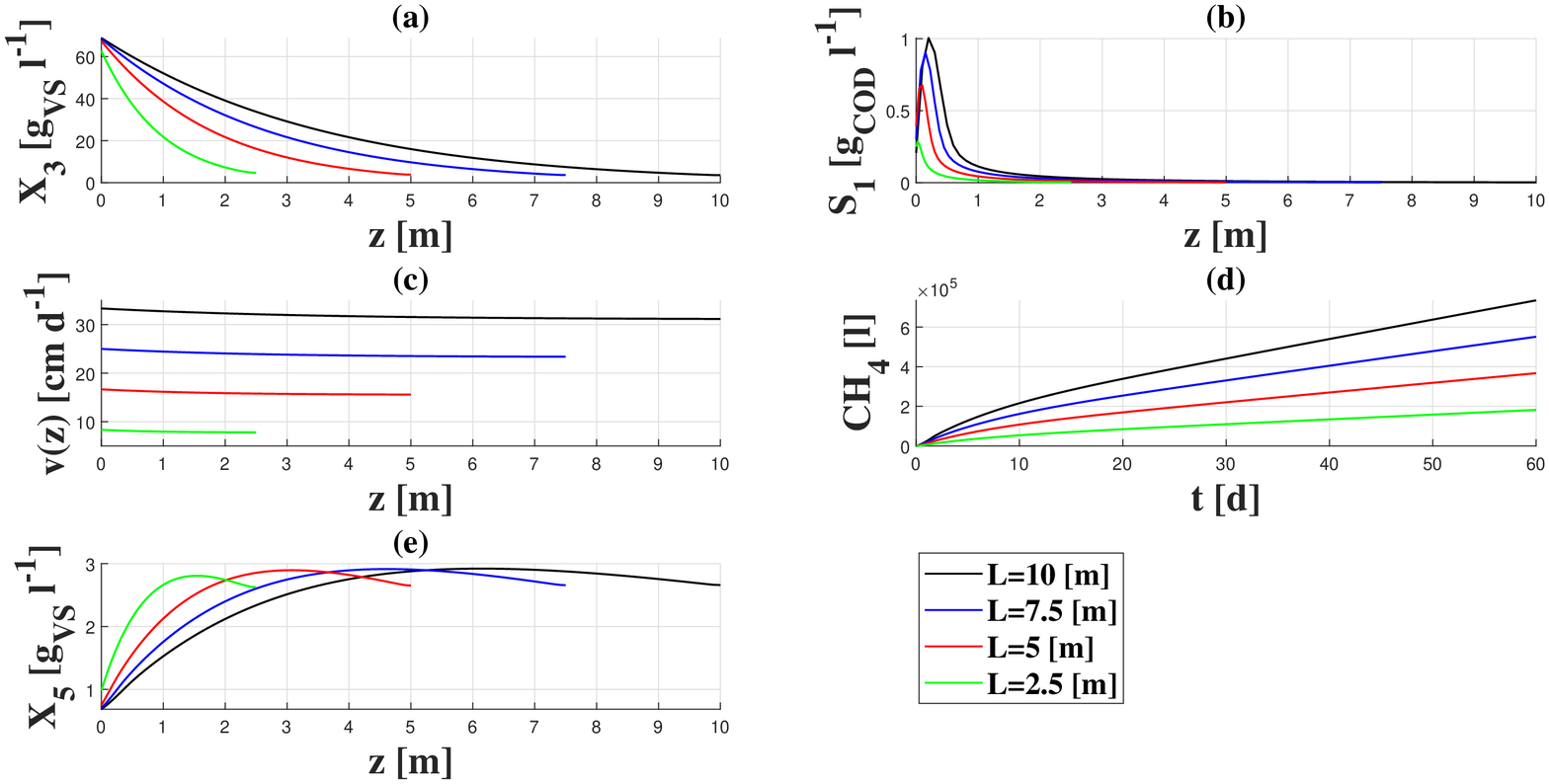}}}
\caption{Bio-degradable VS $X_{3}\left(z,\tau\right)$ (3a), acetic acid $S_{1}\left(z,\tau\right)$ (3b), $v\left(z,\tau\right)$ (3c), methane production (3d) and microbial biomass $X_{5}\left(z,\tau\right)$ (3e) trends for simulations set $A$, at $\tau=\SI{60}{\day}$.}
\label{fig3}
\end{figure}

Observing the results of the simulation set $A$ (Figure \ref{fig3}), it can be noticed that, in this first case, a similar process of consumption of the bio-degradable VS takes place whatever is the reactor length (Figure \ref{fig3}a). This reveals that, treating a different substrate flow rate in reactors with different sizes keeping constant the HRT and OLR can be obtained the same concentration of bio-degradable VS in the outlet flow.

Figure \ref{fig3}b show the acetic acid concentration profiles along $z$ after 60 days of simulation. The acetic acid profile presents a peak which is reached in a position very close to the inlet section for all the tested $L$ values. The acetic acid is then consumed, assuming very low concentration values in the remaining part of the reactor. In this simulation set, where the HRT is kept constant, the values of the acetic acid concentration corresponding to the peaks of the curves decrease as the reactor length decreases.

The function $v(z)$ at simulation time $\tau=\SI{60}{\day}$ is plotted in Figure \ref{fig3}c. The function non-linearly decreases with $z$ according to equation \eqref{eq9} resulting in a 6.5\% lower velocity value in the outlet section for all the tested $L$ values.

The concentrations of the microbial biomass for all the tested cases are reported in Figure \ref{fig3}e. A similar pattern is observed in all cases: in close proximity to the inlet section the concentration of the micro-organisms involved in the consumption of acetic acid is low. Then it increases, assuming its maximum values in the second half of the reactor. Hence, it can be considered that a stratification along the reactor is realized: the processes that lead to the production of acetic acid take place very close to the inlet section; then, thanks to the availability of the nutrients useful for the growth of the micro-organisms acting the uptake of the acetic acid, the acid is consumed and the soluble methane is produced.

Figure \ref{fig3}d shows the methane production in terms of liters in the head-space over time. The results suggest that, for the same period of observation time and under the same HRT and OLR, the longer is the reactor the higher is the methane production.

\begin{figure}
\centering
{\fbox{\includegraphics[scale=0.35]{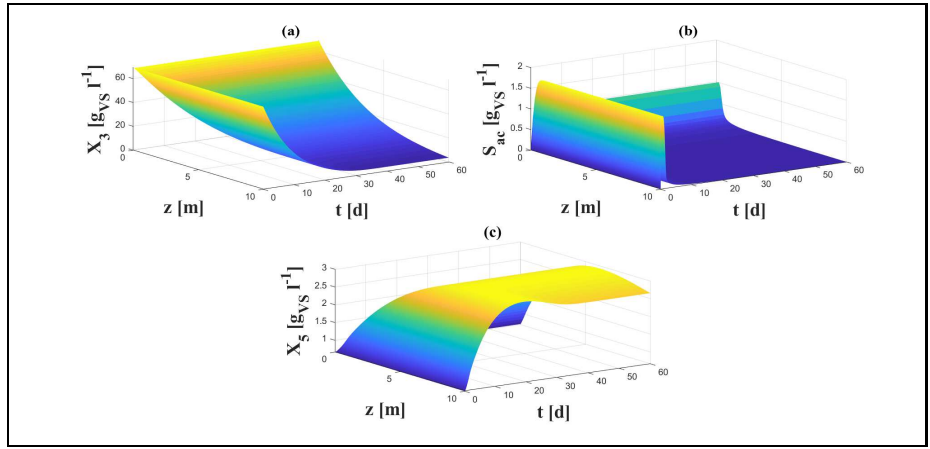}}}
\caption{Bio-degradable VS $X_{3}\left(z,t\right)$ (4a), acetic acid $S_{1}\left(z,t\right)$ (4b) and microbial biomass $X_{5}\left(z,t\right)$ (4c) trends for simulations set $A$, case $A$-4.}
\label{fig3D}
\end{figure}

For the simulation set $A$ are reported also the 3-D plots describing the dynamics of the variables $X_3(z,t)$, $S_1(z,t)$ and $X_5(z,t)$ over time and space in the simulation case $A$-4 (Figure \ref{fig3D}). The constant initial concentration of bio-degradable VS $X_3$ along the reactor is reduced through the degradation process until the profile reported in Figure \ref{fig3}a (case with $L=\SI{10}{\meter}$) is determined. Concerning the acetic acid dynamics it can be observed that it takes place an acid accumulation along the reactor during the initial days but then, with the growth of the concentration of the microbial biomass acting its uptake, the acid is consumed.\\

\renewcommand{\thesubfigure}{\alph{subfigure}}

\subsubsection{Set B} \label{3.2.2}
\begin{figure}
\centering
{\fbox{\includegraphics[width=0.95\textwidth, keepaspectratio]{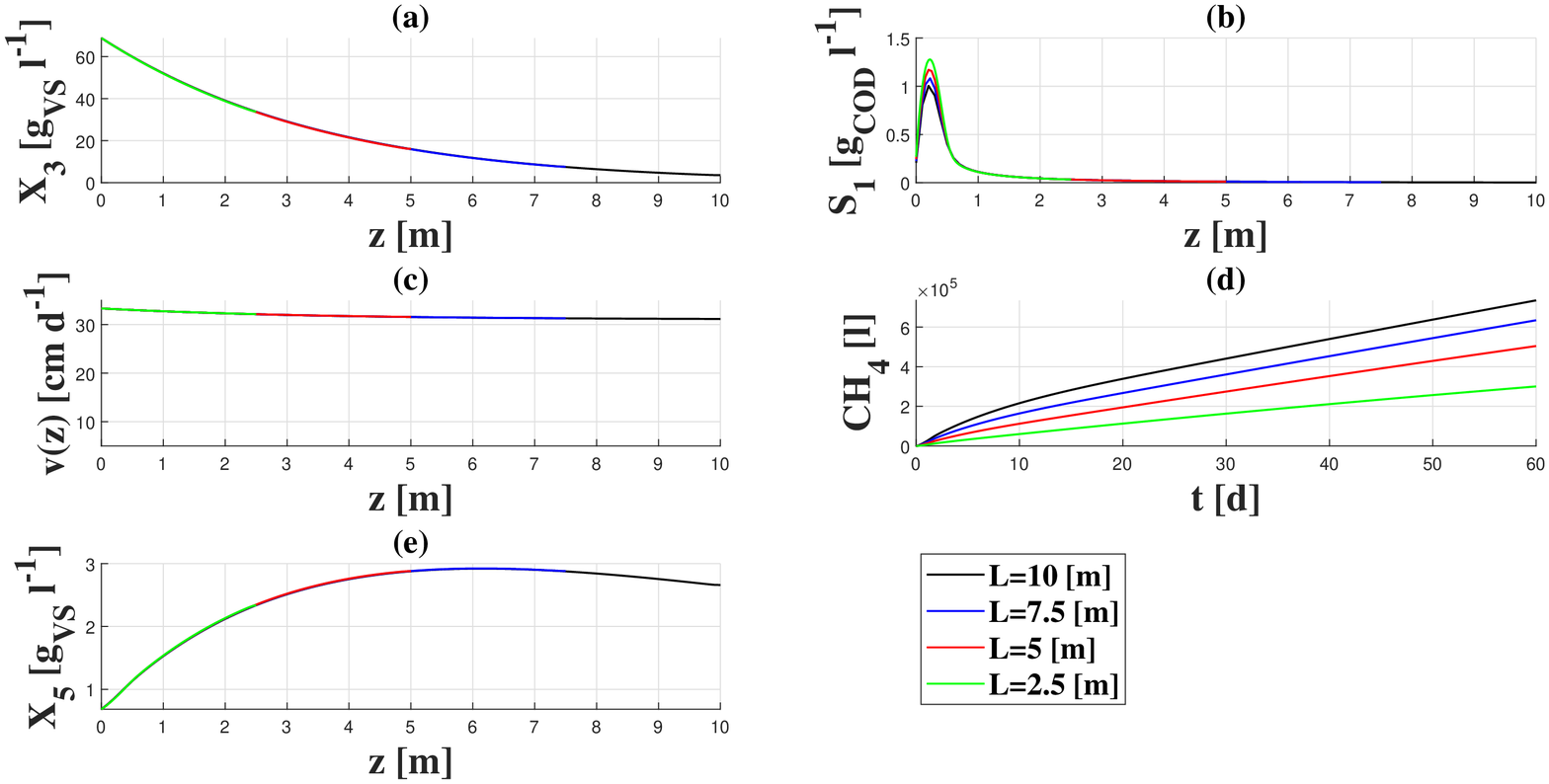}}}
\caption{Bio-degradable VS $X_{3}\left(z,\tau\right)$ (5a), acetic acid $S_{1}\left(z,\tau\right)$ (5b), $v\left(z,\tau\right)$ (5c), methane production (5d) and microbial biomass $X_{5}\left(z,\tau\right)$ (5e) trends for simulations set $B$, at $\tau=\SI{60}{\day}$.}
\label{fig4}
\end{figure}

The results of simulation set $B$ are reported in Figure \ref{fig4}. In this case, the convective inlet velocity $v_0$ of the waste is supposed constant while the reactor length changes. This affects the values of both the HRT and the OLR. First of all, it is possible to observe that all the profiles of the variables along the space are aligned on the same curve, except for the acetic acid. Concerning the bio-degradable VS conversion in acetic acid, it is clear that the shorter is the reactor length the higher is the residual bio-degradable VS concentration (Figure \ref{fig4}a). This is due to the fact that, feeding the waste inside the reactor with a constant inlet velocity, the shorter is the reactor length the lower is the HRT and, using the same concentration of VS, the higher is the OLR. Hence, a growing amount of bio-degradable compounds have less time to be degraded.

The acetic acid profiles show the same patterns established in the case of the simulation set $A$, with peaks located in close proximity to the inlet section and small residues of acid along the remaining part of the reactor. However, in this second case, the values corresponding to the peaks have a small decrease as the reactor length increases.

Moreover, also in the case of the simulation set B the velocity function trends are non-linear with $z$ and it can be noticed a different percentage reduction in the velocity value between the inlet and outlet sections depending on the value of $L$. In particular, the longer is the reactor the higher is this percentage reduction: 3.52\%, 5.24\%, 6.07\% and 6.49\% for the reactor length $L=\SI{2.5}{\meter}$, $L=\SI{5.0}{\meter}$, $L=\SI{7.5}{\meter}$ and $L=\SI{10}{\meter}$ respectively.

The microbial biomass concentration profiles are similar to those observed for the simulation set $A$, with the highest concentration values achieved in the second half of the reactor for each simulation (Figure \ref{fig4}e).

Concerning the liters of produced methane over time (Figure \ref{fig4}d), the results show that, similarly to set $A$, the longer is the reactor the higher is the value of methane produced. This is mostly due to the different consumption of the bio-degradable VS content that takes place among the different operating conditions of the set B. Moreover, the initial mass of VS is higher the longer is the reactor, as a consequence of the initial conditions, and this affects the value of the produced methane. Furthermore, if reactors of the same size of the sets $A$ and $B$ are compared in terms of methane production, the investigated cases of the simulation set $B$ during the same period of simulation time show an higher maximum value of produced methane with respect the cases of set $A$, except for the reactor length $L=\SI{10}{\meter}$ where the operating conditions are the same between the two sets. This reveals that, considering a fixed reactor size, the methane production in absolute terms is improved when an higher volatile solids flow rate is fed into the reactor, even if a lower HRT is adopted, until inhibition phenomena occurs.

\renewcommand{\thesubfigure}{\alph{subfigure}}

\subsubsection{Set C} \label{3.2.3}
\begin{figure}
\centering
{\fbox{\includegraphics[width=0.95\textwidth, keepaspectratio]{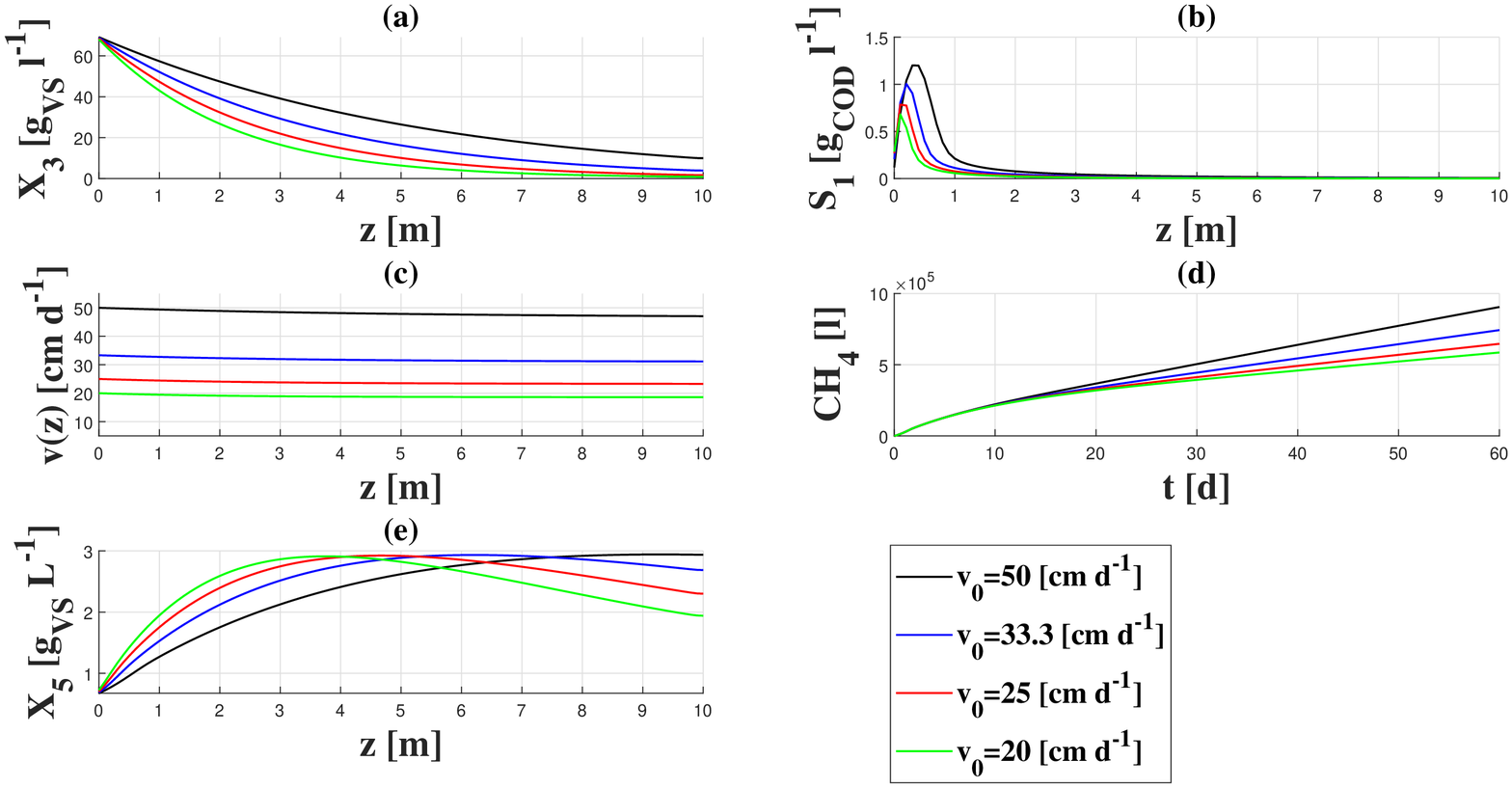}}}
\caption{Bio-degradable VS $X_{3}\left(z,\tau\right)$ (6a), acetic acid $S_{1}\left(z,\tau\right)$ (6b), $v(z)$ (6c), methane production (6d) and microbial biomass $X_{5}\left(z,\tau\right)$ (6e) trends for simulations set $C$, at $\tau=\SI{60}{\day}$.}
\label{fig5}
\end{figure}

The results of simulation set $C$ at simulation time $\tau=\SI{60}{\day}$, as in the previous cases,  are reported in the Figure \ref{fig5}. 

It can be observed that, for all simulations, the removal efficiency of the bio-degradable VS decreases as the inlet convective velocity of the waste increases (Figure \ref{fig5}a). If one analyses 
the removal efficiency of the bio-degradable VS in the three different simulation sets $A$, $B$ and $C$ at the simulation time $\tau=\SI{60}{\day}$ important informations for the designing of new plants of DAD can be obtained: keeping constant the HRT and considering a different reactor lengths and inlet velocities (simulation set $A$) the removal efficiency of the bio-degradable VS does not change so much ($92.5\% \leq \eta \leq 94.8\%$); passing from a length of \SI{10}{\meter} to a length of \SI{2.5}{\meter} (set $B$), keeping constant the inlet velocity of the waste, there is a reduction in the removal efficiency of the bio-degradable VS from $\eta=94.8\%$ to $\eta=50.7\%$; lastly, passing from an inlet convective velocity of the waste of \SI{20}{\centi\meter\per\day} to an inlet velocity of \SI{50}{\centi\meter\per\day} (set $C$), keeping constant the reactor length, there is a reduction in efficiency from $\eta=98.7\%$ to $\eta=85.6\%$. In the latter two cases, the reduction in efficiency means that if the reactor is not long enough or the inlet velocity is too high, respectively, and the process is performed with the objective to reach a certain volatile solids removal, there is the risk to fail.
 
In Figure \ref{fig5}b are reported the results referred to the acetic acid concentration profiles in the case of the simulation set $C$. It can be noticed that the lower is the inlet velocity the lower is the corresponding peak of the acetic acid concentration curve and the nearer to the inlet section this peak is reached. The residual acetic acid concentration show that, also in this simulation set, the acid is completely consumed along the reactor.

The non-linearity characterizes the variation of the convective velocity along the reactor also in all cases of the set $C$ (Figure \ref{fig5}c). Moreover, the curves describing the concentration of the micro-organisms involved in the consumption of acetic acid profile along the reactor length reported in Figure \ref{fig5}e present a visible decreasing pattern developed in the second half of the reactor, mainly in the cases where the inlet velocity is lower. These trends are the consequence of the differences in  the convective transport phenomena and the acetic acid consumption dynamics occurring among the analysed cases.

Concerning the liters of methane in the head-space over time (Figure \ref{fig5}d), results confirm that, once fixed the reactor length, the higher is the adopted inlet velocity of waste fed into the reactor, which means also an higher OLR, the higher is the maximum value of produced methane. 
\renewcommand{\thesubfigure}{\alph{subfigure}}

\begin{figure}
\begin{framed}
\centering
\subfloat[][]
{\fbox{\includegraphics[width=0.7\textwidth, keepaspectratio]{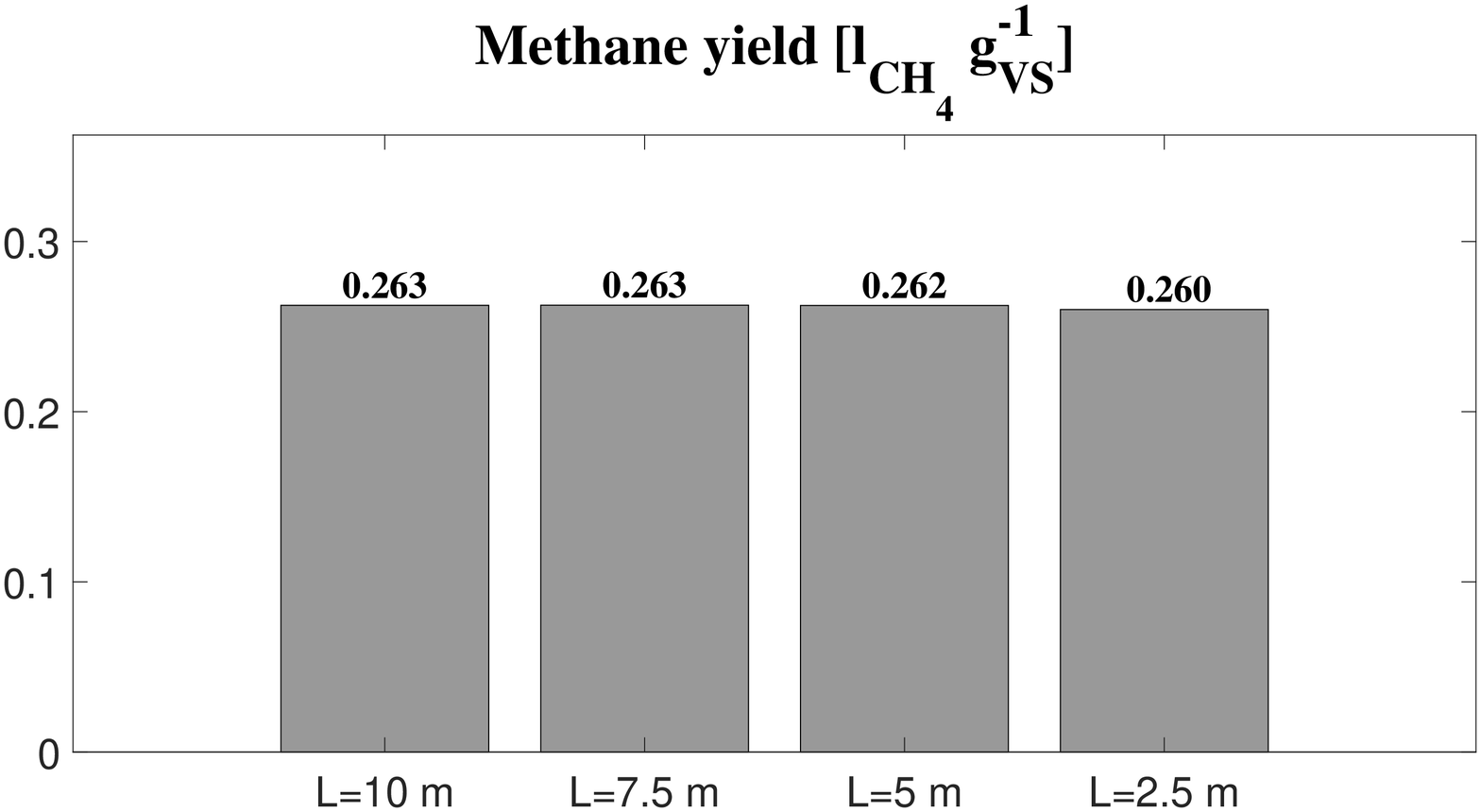}}\label{subfig6.1}} \hspace{0.1cm}
\subfloat[][]
{\fbox{\includegraphics[width=0.7\textwidth, keepaspectratio]{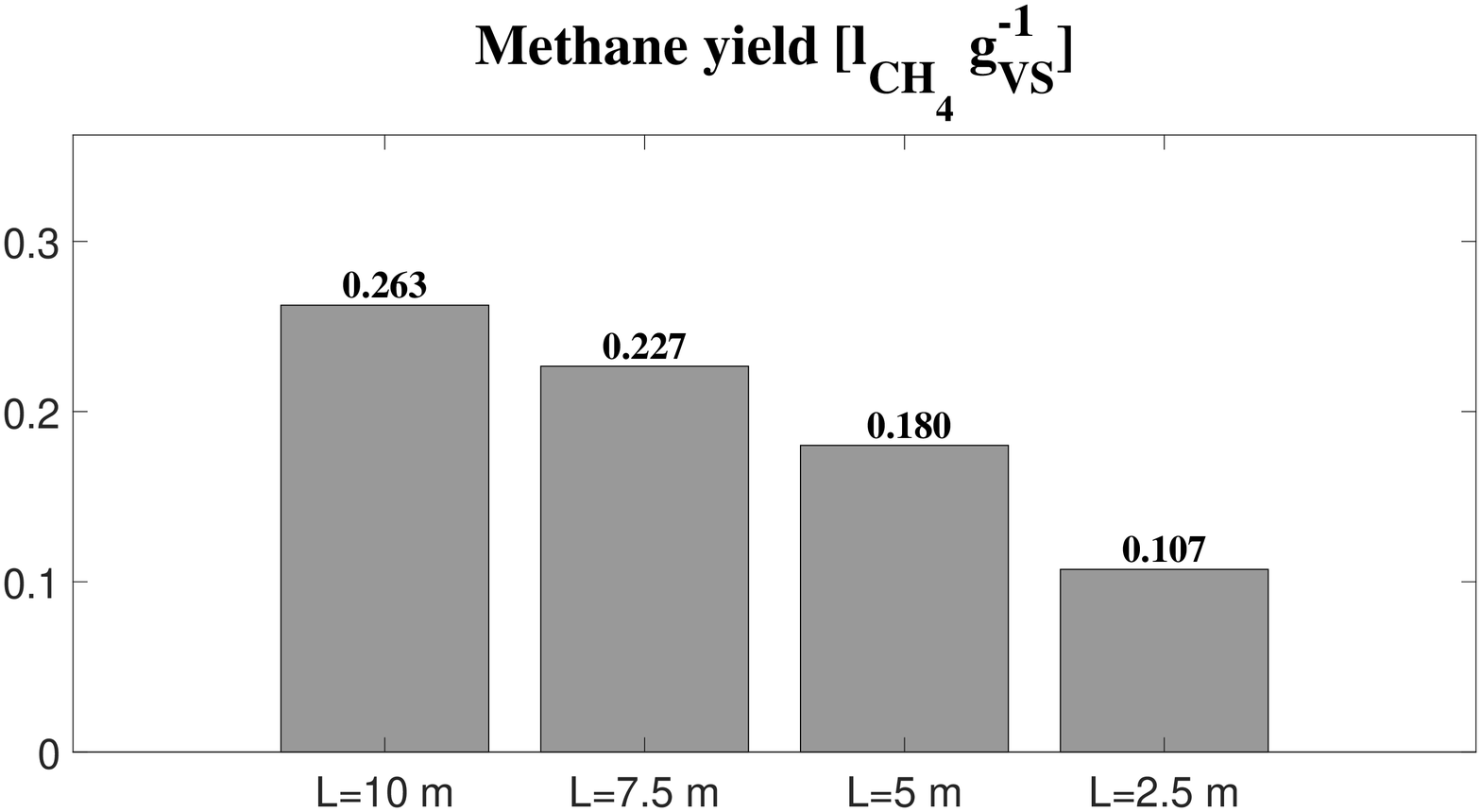}}\label{subfig6.2}} \\
\subfloat[][]
{\fbox{\includegraphics[width=0.7\textwidth, keepaspectratio]{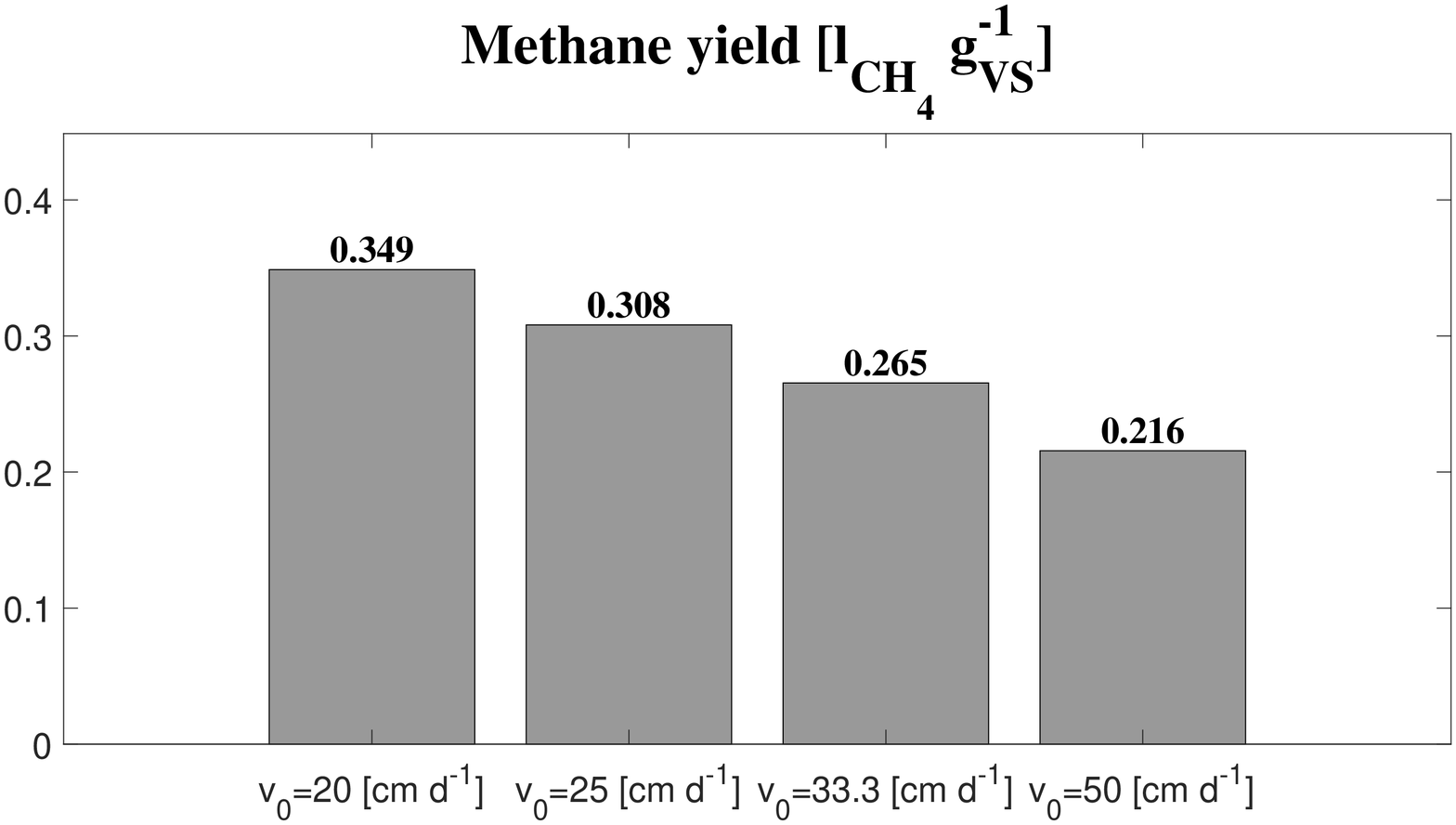}}\label{subfig6.3}} \\
\end{framed}
\caption{Methane yield for simulations sets $A$ (\ref{subfig6.1}), $B$ (\ref{subfig6.2}) and $C$ (\ref{subfig6.3}) after $\tau=\SI{60}{\day}$}
\label{fig6}
\end{figure}

However, the results of methane production of the three simulation sets $A$, $B$ and $C$ can be analysed also monitoring the methane yields $y$, evaluated as the ratio between the value of the methane production curve at the instant time $\tau=\SI{60}{\day}$ $\alpha$ $\left[ l_{CH_4} \right]$ over the added mass of VS $\beta$ $\left[ g_{VS}\cdot \right]$ in the time interval $\left[ 0,\tau \right]$  (Eq. \eqref{eq25}) and reported in Figure \ref{fig6}. 
\begin{equation}
y=\frac{\alpha(\tau)}{\beta(\tau)} \label{eq25}
\end{equation} 
It is possible to observe that in the case of simulation set $A$ the yields are constant with the reactor length (Figure \ref{subfig6.1}), while in the simulation set $B$ the yield is about 2.5 times higher going from $L=\SI{2.5}{\metre}$ to $L=\SI{10.0}{\metre}$ (Figure \ref{subfig6.2}). Moreover, comparing the methane yields of the simulation sets $A$ and $B$ it can be concluded that, despite the absolute methane production increases when reactors of the same length are fed with an higher amount of volatile solids, the yield decreases. A similar result is observed in simulation set $C$. Indeed, in this set is analysed the methane production of reactors having the same size fed using an increasing inlet velocity of the waste and an increasing OLR and, contrary to the maximum value of produced methane which increases as the adopted inlet velocity of waste increases, the methane yield decreases (Figure \ref{subfig6.3}).

\renewcommand{\thesubfigure}{\alph{subfigure}}

\begin{figure}
\begin{framed}
\centering
\subfloat[][Bio-degradable VS $X_{3}\left(z,\tau\right)$ concentration trend.]
{\fbox{\includegraphics[width=0.95\textwidth, keepaspectratio]{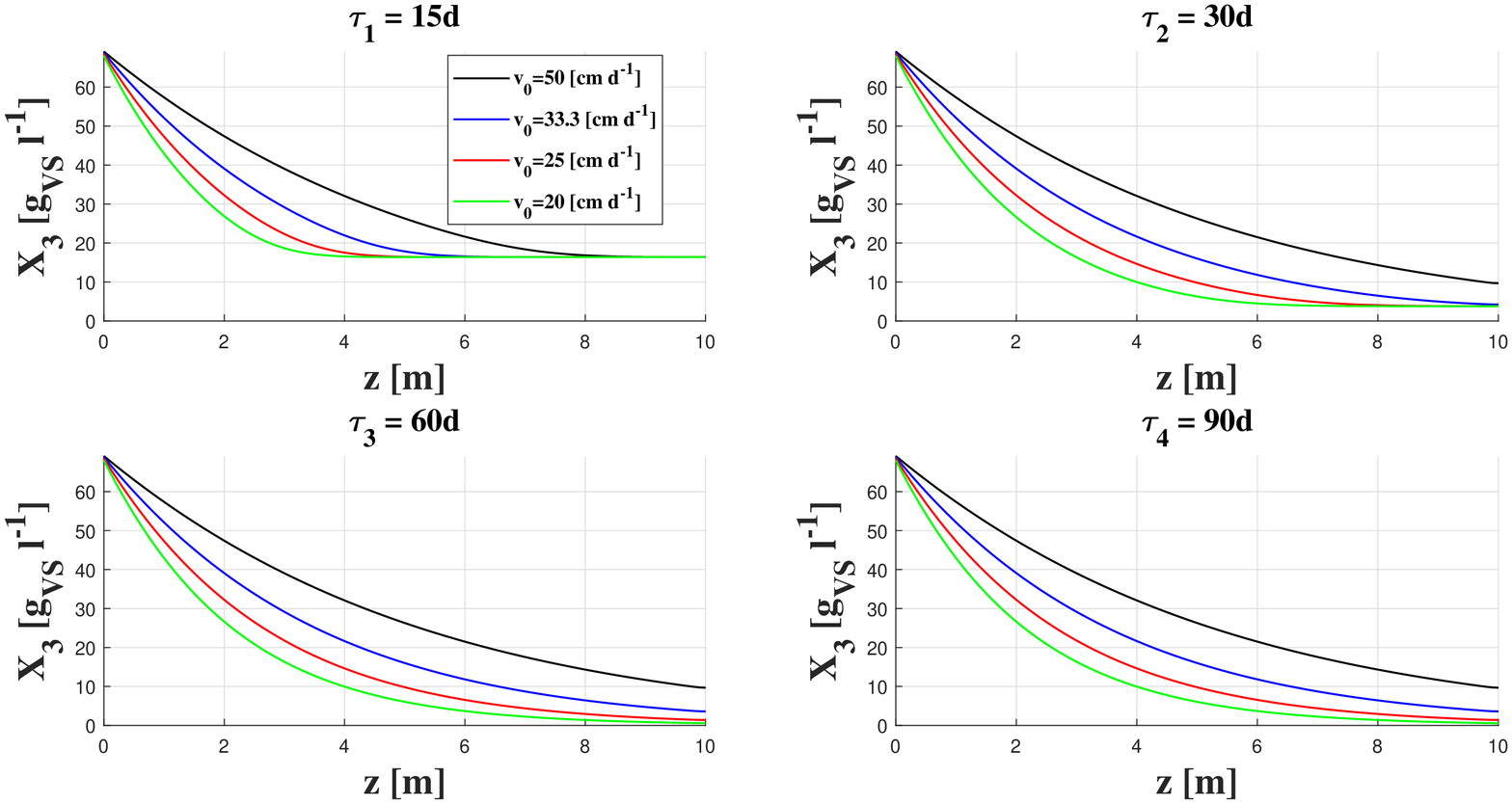}}\label{subfig7.1}} \\
\subfloat[][Microbial biomass $X_{5}\left(z,\tau\right)$ concentration trend.]
{\fbox{\includegraphics[width=0.95\textwidth, keepaspectratio]{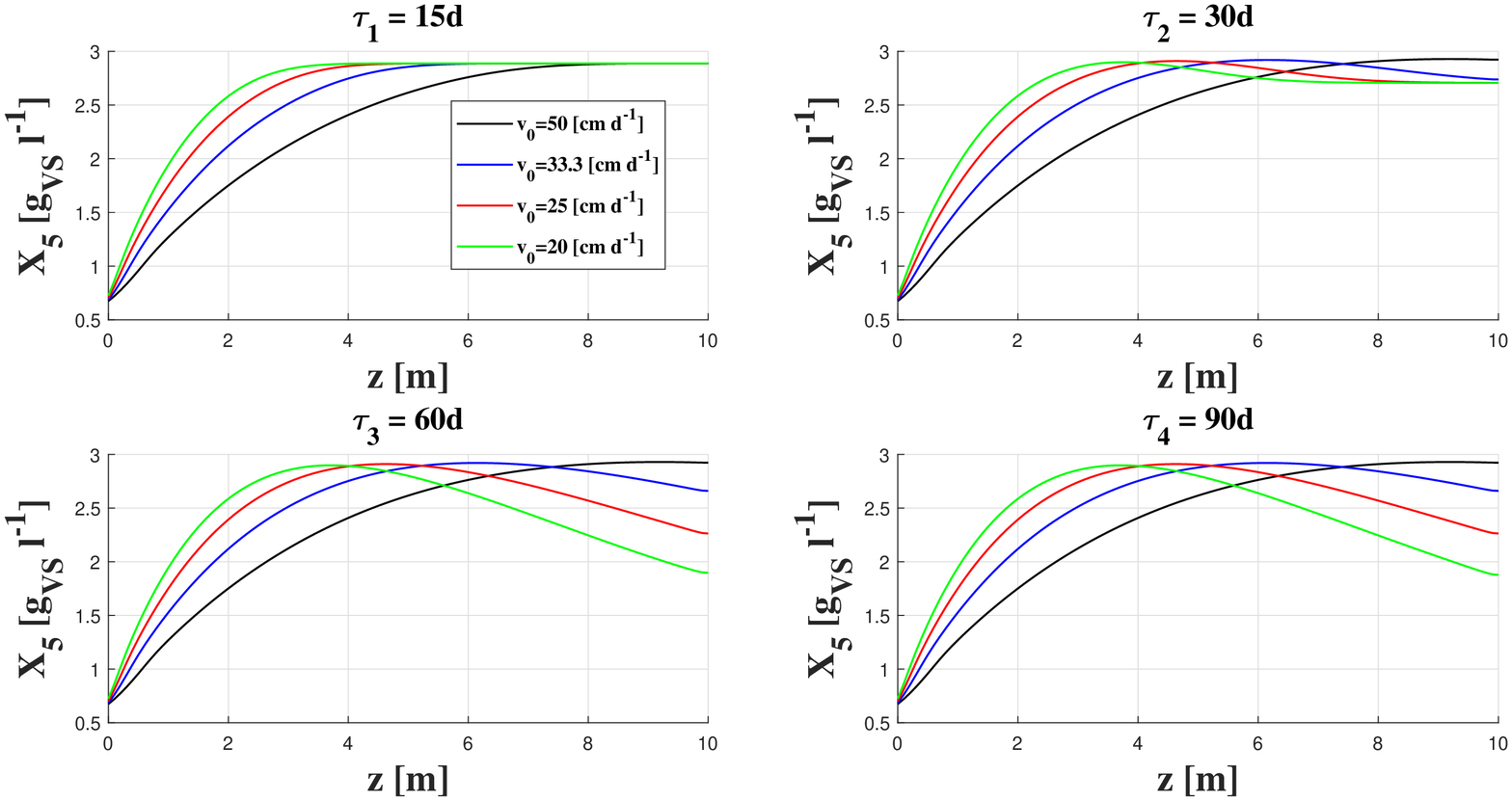}}\label{subfig7.2}} \\
\end{framed}
\caption{Bio-degradable VS $X_{3}\left(z,\tau\right)$ (\ref{subfig7.1}) and microbial biomass $X_{5}\left(z,\tau\right)$ (\ref{subfig7.2}) concentration trends at different time of simulation $\tau$ when $L=cost$ and $v_0$ varies, simulations set $C$.}
\label{fig7}
\end{figure}

\begin{figure}
\begin{framed}
\centering
\subfloat[][Acetic acid $S_{1}\left(z,\tau\right)$ concentration trend.]
{\fbox{\includegraphics[width=0.95\textwidth, keepaspectratio]{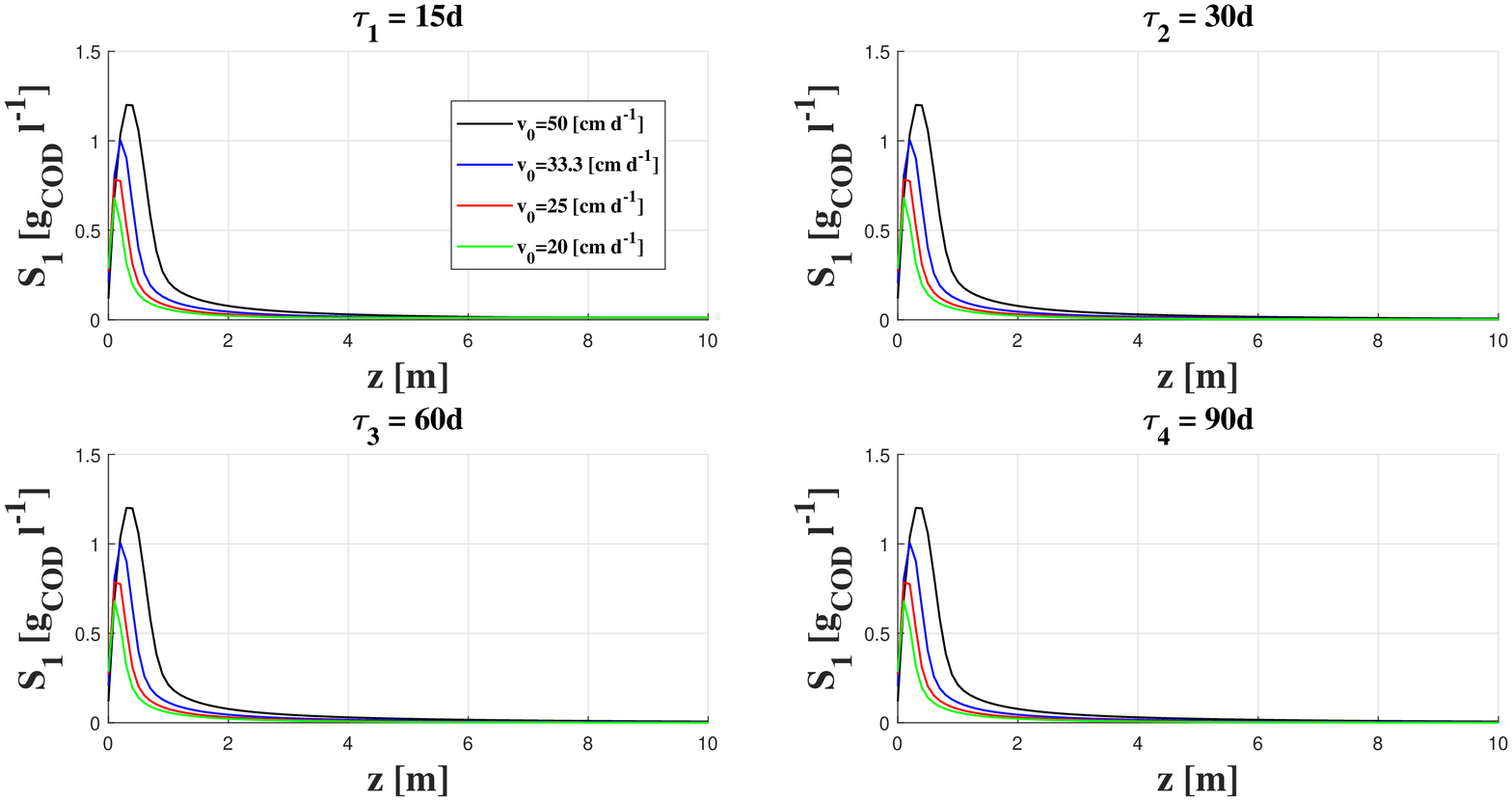}}\label{subfig8.1}}
\\
\subfloat[][$v\left(z,\tau\right)$ trend.]
{\fbox{\includegraphics[width=0.95\textwidth, keepaspectratio]{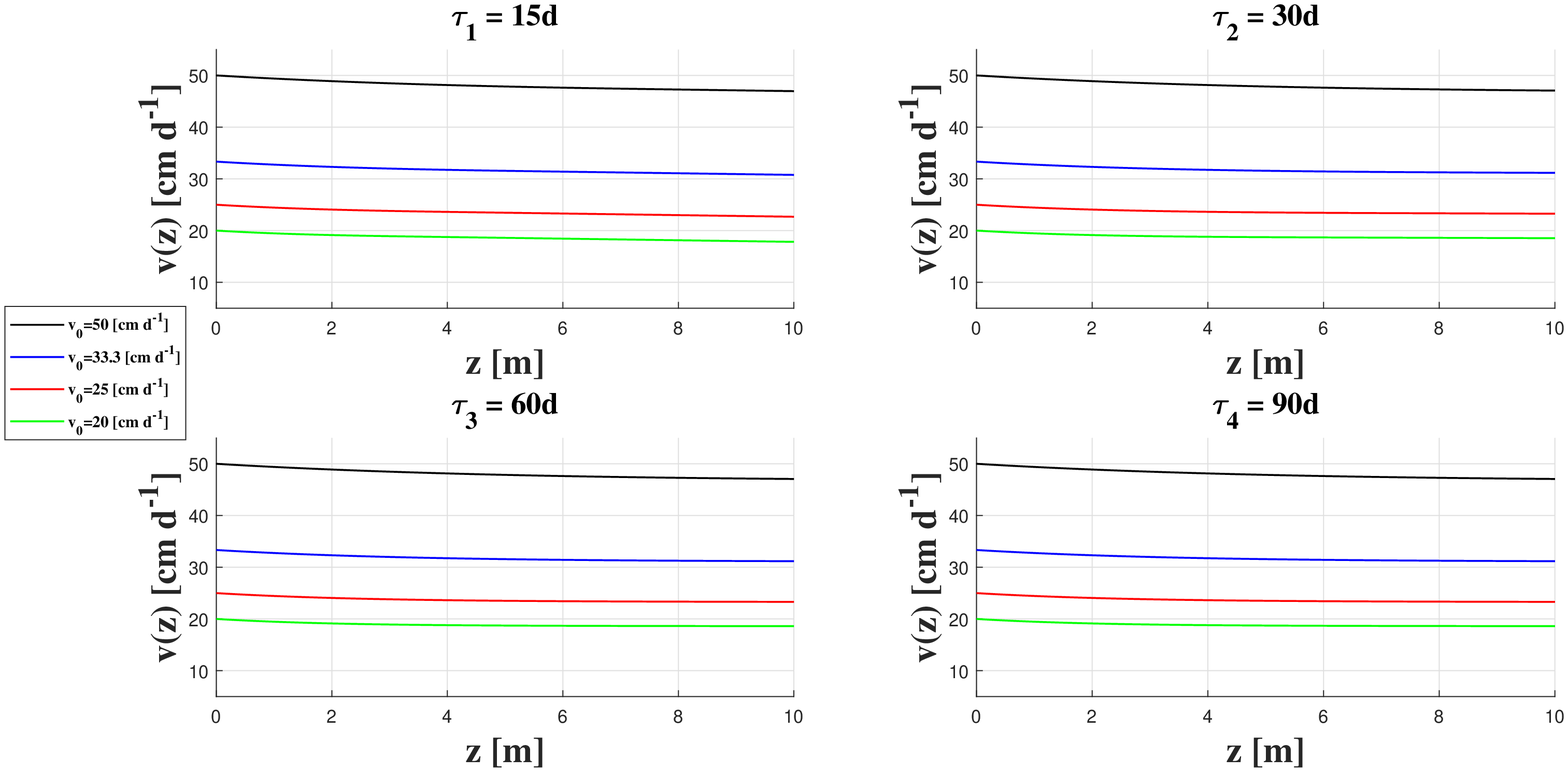}}\label{subfig8.2}} \\
\end{framed}
\caption{Acetic acid concentration $S_{1}\left(z,\tau\right)$ (\ref{subfig8.1}) and $v\left(z,\tau\right)$ (\ref{subfig8.2}) trends at different time of simulation when $L=cost$ and $v_0$ varies, simulations set $C$.}
\label{fig8}
\end{figure}

Furthermore, for the simulation set $C$ the bio-degradable VS concentration $X_{3}\left(z,\tau\right)$, microbial biomass concentration $X_{5}\left(z,\tau\right)$, acetic acid concentration $S_{1}\left(z,\tau\right)$ trends and $v\left(z,\tau\right)$ profile at simulations times $\tau_1=\SI{15}{\day}$, $\tau_2=\SI{30}{\day}$, $\tau_3=\SI{60}{\day}$ and $\tau_4=\SI{90}{\day}$ are reported in Figures \ref{fig7} and \ref{fig8}. Concerning the dynamics of the bio-degradable VS and microbial biomass concentrations reported in Figures \ref{subfig7.1} and \ref{subfig7.2} respectively, there are no differences between the 60th and 90th day, revealing that the process reached the steady state in all the investigated cases. On the contrary, the acetic acid concentration and velocity profiles along $z$ appear to reach the steady state within the first 15 days (Figure \ref{subfig8.1} and \ref{subfig8.2}).

\renewcommand{\thesubfigure}{\alph{subfigure}}

\subsubsection{Set D} \label{3.2.4}
\begin{figure}
\centering
{\fbox{\includegraphics[width=0.95\textwidth, keepaspectratio]{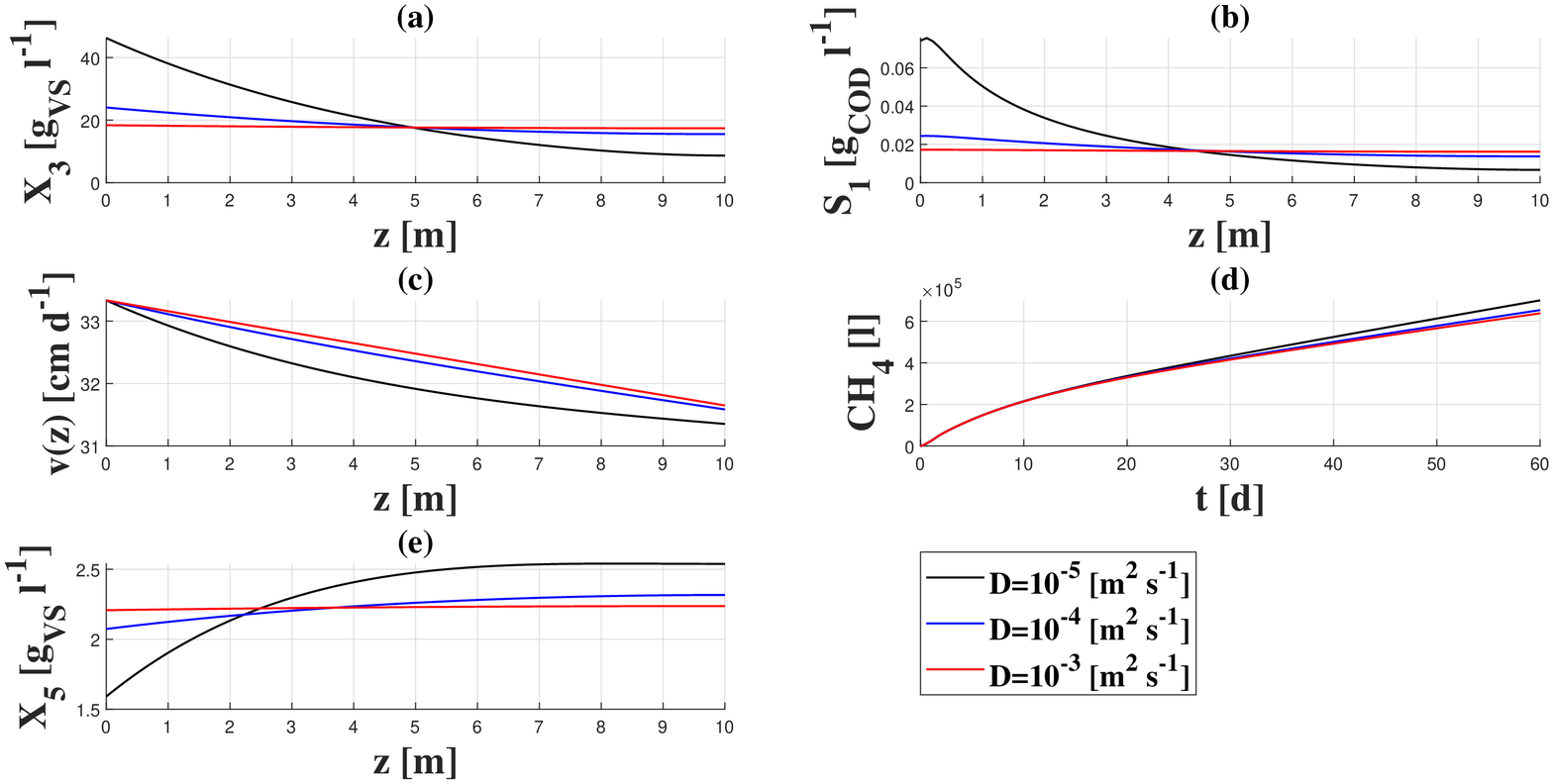}}}
\caption{Bio-degradable VS $X_{3}\left(z,\tau\right)$ (10a), acetic acid $S_{1}\left(z,\tau\right)$ (10b), $v\left(z,\tau\right)$ (10c), methane production (10d) and microbial biomass $X_{5}\left(z,\tau\right)$ (10e) trends for simulations set $D$, at $\tau=\SI{60}{\day}$.}
\label{fig10}
\end{figure}

Figure \ref{fig10} shows the results of the simulation set $D$, where the effects of the diffusion coefficient $\bar{D}$ are investigated, keeping constant the other operating and physical parameters. Based on the simulation results, it is possible to state that the diffusion coefficient strongly affects the reactor performances.
In fact, passing from an order of magnitude of $10^{-5}$ to $10^{-3}$ it is clear that diffusion contributes to the homogenization of the compounds along the reactor. Among the consequences there is the fact that, when the diffusion coefficient is higher, the velocity function assumes a linear trend with $z$, due to the constant concentrations along the reactor axis of the variables on which it depends. Moreover, it can be noticed that the homogenization of compounds inside the reactor explicitly show the difference between a CSTR and a PFR: there is a lower average concentration of substrates inside the reactor when the homogenization is high (CSTR behaviour), and this leads to have a lower uptaking rate for the substrates. As a consequence, the methane production is maximized the lower is the diffusion coefficient value (Figure \ref{fig10}d).\\

\renewcommand{\thesubfigure}{\alph{subfigure}}

All these behaviours observed through the previous simulations are in accordance with the physics of the phenomenon, showing that the model is capable to correctly represent process dynamics. These informations could be used for the designing of anaerobic digestion in PFR plants. The optimal length or inlet velocity could be chosen depending on the result to be maximized: for example, the choice of a longer or shorter reactor depends on whether it is important to maximize the methane production or the organic waste removal using a fixed value of the HRT. Furthermore, monitoring the acid concentrations inside the reactor could be crucial to avoid system failure. Lastly, the system velocity could be predicted gaining information useful to establish the outlet flow in order to avoid reactor emptying, which could cause system locking.

\subsection{Application to a real case} \label{n3.3}
To show model consistency with the bio-physics of a DAD process, the present model was used to simulate the dynamics of a laboratory scale digester. The reference experimental work is that presented by Regina J. Patinvoh et al. \cite{patinvoh2017dry}, who performed a DAD process of untreated manure bedded with straw at 22\% of TS content under different OLRs. The experimental campaign started with a process of AD in batch conditions, established for reactor initialization. At this step, a mixture of inoculum and substrate in a ratio of 2:1 in terms of volatile solids content, was used as reactor feeding. This batch phase lasted 40 days and later the reactor was fed continuously according to the feeding conditions of Table \ref{table8}. Firstly, the $OLR=2.8 \ \si{\gram\per\liter\per\day}$ was used and then it was increased gradually to 4.2 and 6.0 $\si{\gram\per\liter\per\day}$. The corresponding decreasing HRTs were of 60, 40 and 28 days. The feeding conditions were changed after a time period equal to the HRTs. Using the characteristics of the substrate and of the inoculum, reported in Table \ref{table7} and reproducing the feeding procedure of \cite{patinvoh2017dry} results concerning daily methane production, cumulative biogas production and VFA concentration path were reproduced. The used reactor is an horizontal plug-flow with a total volume of \SI{9.2}{\L}; the temperature regime is set to \SI{37}{\celsius}. 
The values of model parameters used for reproducing experimental data are the same as reported in Table \ref{table1}, except for the kinetic constant for the consumption of the volatile solids $k_1$, which was set equal to $0.035 \ d^{-1}$. The values of the conversion coefficient of VS in COD $m$ and the fraction of decayed microbial biomass becoming new bio-degradable substrate $f$ were set as in the previous simulations.\\

Since no informations were available concerning the initial microbial biomass concentration, preliminary simulations of the batch case were run to build the missing initial condition. Later, the batch conditions of the experimental campaign were simulated, this time with the aim to reproduce the methane production using the built initial condition on the microbial biomass concentration. The values of the simulated variables at the end of this period were used as initial conditions for the simulation of the feeding condition 1. 

\begin{table}
\centering
\begin{threeparttable}
\footnotesize
\begin{tablenotes}
\item $ND=not \ determined.$
\end{tablenotes}
\makebox[\linewidth]{
\begin{tabular}{lcccc}
\toprule
\multicolumn{1}{c}{}  & \textbf{Symbol}      & \textbf{Unit}       & \textbf{Manure with straw} & \textbf{Inoculum} \\ \midrule
Moisture              & $f_1$                & $g/g$            & 0.7772                     & 0.9220             \\
TS content            & $\left(1-f_1\right)$ & $g_{TS}/g$        & 0.2229                       & 0.078              \\
VS content on TS base & $f_2$                & $g_{VS}/g_{TS}$   & 0.7044                     & 0.4046             \\
Ash                   & $\left(1-f_2\right)$ & $g/g_{TS}$        & 0.2956                     & 0.5954 \\
$COD_{content}$      &                      & $g_{COD}/g_{VS}$  & 0.73                       & ND                 \\
$BMP_{theoretical}$  &                      & $L_{CH_4}/g_{VS}$ & 0.290                       & ND               \\
\bottomrule
\end{tabular}}
\caption{Substrate and Inoculum characteristics used during the experimental campaign of \cite{patinvoh2017dry}.}
\label{table7}
\end{threeparttable}
\end{table}

\begin{table}
\begin{small}
\begin{center}
\renewcommand\arraystretch{1.3}
\begin{tabular}{lcccc}
\toprule
{\textbf{Parameter}} & {\textbf{Unit}} & \multicolumn{3}{c}{\textbf{Condition}} \\
\multicolumn{2}{c}{} & 1 & 2 & 3  \\
\midrule
OLR    & $g_{VS}\ L^{-1}\ d^{-1}$  &2.8 &  4.2                    & 6.0  \\
HRT     & $d$ &60 &  40 & 28  \\
Loading rate & $g_{VS}\ d^{-1}$ & 13.8 & 20.7 & 29.9 \\
\bottomrule
\end{tabular}
\caption{OLRs, HRTs and Loading rates used in the experimental work of Patinvoh et al. \cite{patinvoh2017dry} and in the simulations used to reproduce its results.} \label{table8}
\end{center}
\end{small}
\end{table}

\begin{figure}
\centering
{\fbox{\includegraphics[width=0.95\textwidth, keepaspectratio]{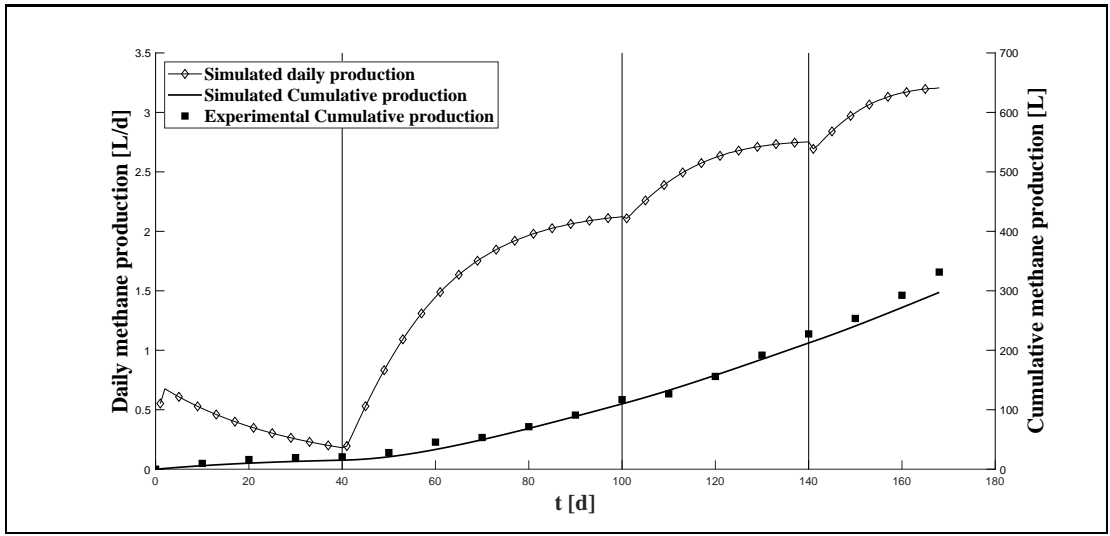}}} 
\caption{Daily and cumulative methane production in the experimental and simulated cases.}
\label{fig11}
\end{figure}

The resulting dynamics determined the initial conditions for the feeding conditions 2 and so on.\\
It is possible to observe that the model results reasonably follow the experimental data (Figure \ref{fig11}).

In order to compare the experimental and simulated methane productions the experimental cumulative biogas production curve reported in \cite{patinvoh2017dry} has been multiplied by the measured average methane content. The experimental daily methane production curve is not reported here, but to show that the simulated daily production well reproduces the experimental results, the simulated and experimental methane yields have been compared.

\begin{table}
\begin{small}
\begin{center}
\renewcommand\arraystretch{1.3}
\begin{tabular}{lcccc}
\toprule
{\textbf{Feeding condition}} & \multicolumn{2}{c}{\textbf{Methane yield}} \\ & \multicolumn{2}{c}{$\left[ L_{CH_4}\ g_{VS}^{-1} \right]$}   \\
& Experimental & Simulated  \\
\midrule
OLR 1    &  0.16     & 0.16  \\
OLR 2    &  0.17      & 0.14   \\
OLR 3    &  0.14      & 0.12   \\
\bottomrule
\end{tabular}
\caption{Experimental and Simulated methane yield.} \label{table10}
\end{center}
\end{small}
\end{table}

Table \ref{table10} show that the model is very good in predicting the system evolution. Similarly to the experimental case, the simulated daily methane production is decreasing during the development of the process in batch conditions. In the continuous feeding condition cases the profiles are increasing until a certain maximum value and then remain constant. The VFA path is well followed by the model. As in the experimental results, simulated acid concentration in the last section of the reactor is very low during all the simulation time (data not shown).  
\renewcommand{\thesubfigure}{\alph{subfigure}}
\section{Conclusions} \label{n4}
In this work a mathematical model for the dry anaerobic digestion process in PFRs based on the convection-diffusion-reaction equation in one-dimensional case was presented. The model predicts substances concentration along the reactor and it is also capable to describe the variation of the system velocity with time and space. The equation governing the convective velocity is derived by considering the hypothesis that density of the waste matrix within the reactor is constant over time and the sum of the volume fractions of the bio-components constituting the waste matrix are constrained to sum up to unity. The system of PDEs was integrated numerically. Results show model consistency with experimental evidence at laboratory scale and highlight the importance to have a mathematical tool useful to manage, size and improve real plants.
For future development the kinetic model may be extended to the complete ADM1 framework. This will allow to predict all species dynamics, such as volatile fatty acids concentration profiles that can be used to predict the pH profile along the reactor and take into account inhibition processes on system dynamics. Moreover the composition of the biogas could be characterized, allowing to determine its effective energy power.
Additionally, the equation describing methane concentration in the head-space may be modified by introducing a loss term that reproduces a certain gas tapping from the head-space of the reactor, bringing the model closer to the operative conditions of real plants. Another improvement of the model could consist in considering a variable density of the treated matrix. Kinetic processes could be linked to diffusion and the model could also be extended to 2D and 3D domains. This could allow to analyse other phenomena occurring along directions different from the system movement direction that may affect the anaerobic digestion process performance. Moreover, a sensitivity analysis is needed in order to investigate the most influencing parameters of the model and a qualitative analysis of the solutions is needed to prove existence, uniqueness and stability of the solutions.


\section{Acknowledgements}
This work has been developed in the context of D.D. n. 1377 on June 5, 2017, additional PhD fellowships for 2017/2018 academic year, course XXXIII within the framework of the ”Programma Operativo Nazionale Ricerca e Innovazione (PON RI 2014/2020) Action I.1 - Innovative PhDs with industrial characterization”. Research project: MATHEMATICAL MODELING OF DRY ANAEROBIC DIGESTION.\\

This paper has been performed under the auspices of the G.N.F.M. of I.N.d.A.M.

\section*{Nomenclature}
\begin{longtable}{ll}
$HRT$    &    Hydraulic Retention Time $\left( d \right)$ \\
$OLR$    &    Organic Loading Rate  $\left( g_{VS}\ l^{-1}\              											d^{-1} \right)$  \\
$\alpha$ &  Cumulative methane production value at a certain instant time  $\left( l_{CH_4}											  \right)$  \\
$\beta$  & Added mass of volatile solids in a certain time interval $\left( g_{VS} \right)$ \\
$\rho$ & Waste density $\left( kg \ m^{-3} \right)$\\

$\tau$ & Simulation time $\left( d \right)$ \\

$A_{w}$ & Cross-section of the reactor volume occupied by waste $\left( m^{2} \right)$ \\

$\bar{D}$ & Diffusion coefficient $\left( m^2 \ s^{-1} \right)$ \\

$F_i$ & Source/Consumption term of the $i^{th}$ component $\left( kg \ m^{-3} \ s^{-1} \right)$ \\

$g$ & Mass flux per unit area $\left( kg \ m^{-2} \ s^{-1} \right)$ \\

$G$ & Gaseous methane concentration $\left( kg_{COD} \ m^{-3} \right)$  \\

$k_1$ & Kinetic constant for the consumption of the volatile solids $\left( d^{-1} \right)$ \\

$k_2$ & Monod maximum specific uptake rate $\left( d^{-1} \right)$\\

$k_3$ & Gas-liquid transfer coefficient $\left( d^{-1} \right)$ \\

$k_4$ & First order decay rate of the biomass $\left( d^{-1} \right)$ \\
 
$K_1$ & Half saturation constant $\left( kg_{COD} \ m^{-3}  \right)$\\

$K_H$ & Henry's law coefficient $\left(M bar^{-1}\right)$ \\

$L$ & Reactor length $\left( m \right)$  \\

$m$ & Conversion factor of volatile solids in COD $\left( kg_{COD}\ kg_{VS}^{-1} \right)$  \\

$r_1$ & Kinetic rate for the process of consumption of volatile solids $\left( kg_{VS}\ m^{-3}\ d^{-1} \right)$ \\

$r_2$ & Methanogenesis kinetic rate $\left( kg_{VS}\ m^{-3}\  d^{-1} \right)$ \\

$r_3$ & Gas-transfer kinetic rate $\left( kg_{COD}\ m^{-3}\  d^{-1} \right)$ \\

$r_4$ & Death rate of the microbial biomass $\left( kg_{VS}\ m^{-3}\  d^{-1} \right)$ \\

$R$ & Gas law constant $\left(bar M^{-1} K^{-1} \right)$ \\

$S_1$ & Soluble acetic acid concentration $\left( kg_{COD} \ 													m^{-3} \right)$ \\
$S_{2}$ & Soluble methane concentration $\left( kg_{COD} \ 													m^{-3} \right)$ \\

$T$ & Temperature $\left( K \right)$ \\

$v$ & Velocity of the compounds moving along the reactor $\left( m \ s^{-1} \right)$ \\

$v_0$ & Waste inlet velocity $\left( m \ s^{-1} \right)$ \\

$V_{gas}$ & Head-space volume for the gas storage $\left( m^{3} \right)$ \\

$X_1$ & Water concentration $\left( kg \ m^{-3} \right)$  															\\
$X_2$ & Inert concentration $\left( kg \ m^{-3} \right)$ \\
$X_3$ & Bio-degradable VS concentration $\left( kg_{VS} \ 													m^{-3} \right)$ \\
$X_4$ & Non Bio-degradable VS concentration $\left( kg_{VS} \ 												m^{-3} \right)$ \\
$X_{5}$& Microbial biomass acting the uptake of acetic acid concentration 						$\left( kg_{VS} \ m^{-3} \right)$ \\

$y$ & Methane yield $\left( l_{CH_4} \ g_{VS}^{-1} \right)$ \\

$Y$ & Yield of biomass on substrate $\left( - \right)$

\end{longtable}
\bibliographystyle{unsrt}
\bibliography{Plug_Flow_Arxiv}

\Addresses

\end{document}